\documentclass[aip,amsmath]{revtex4-2}

\usepackage{pstricks,graphicx,bbm,amssymb,booktabs}
\usepackage[utf8]{inputenc}
\usepackage{hyperref}
\usepackage[english]{babel}
\usepackage{color}

\def\Id{{\mathbbm 1}}

\def\bflambda{\boldsymbol{\lambda}}
\def\blambda{\boldsymbol{\lambda}}
\def\Tr{{\rm Tr}}
\def\FI{{\boldsymbol{\cal F}}}
\def\QFI{{\boldsymbol{\cal H}}}
\def\F{{\cal F}}
\def\H{{\cal H}}
\def\UHL{\boldsymbol{\cal U}}
\def\p{{\rm p}}
\def\a{{\rm a}}
\def\b{{\rm b}}
\def\c{{\rm c}}
\def\caseI{{\rm I}}
\def\caseII{{\rm II}}
\def\HD{{\rm h}}
\def\HET{{\rm dh}}

\newcommand{\MN}[1]{{\color{black}#1}}

\begin{document}
\title[]{\large Joint estimation of noise and nonlinearity in Kerr systems}

\author{Michele~N.~Notarnicola} 
\affiliation{Dipartimento di Fisica ``Aldo Pontremoli'',
Universit\`a degli Studi di Milano via Celoria 16, I-20133 Milano, Italy}
\address{INFN, Sezione di Milano, via Celoria 16, I-20133 Milano, Italy}
\author{Stefano Olivares}
\affiliation{Dipartimento di Fisica ``Aldo Pontremoli'',
Universit\`a degli Studi di Milano via Celoria 16, I-20133 Milano, Italy}
\address{INFN, Sezione di Milano, via Celoria 16, I-20133 Milano, Italy}
\author{Matteo~G.~A. Paris}\email{matteo.paris@fisica.unimi.it}
\affiliation{Dipartimento di Fisica ``Aldo Pontremoli'',
Universit\`a degli Studi di Milano via Celoria 16, I-20133 Milano, Italy}
\address{INFN, Sezione di Milano, via Celoria 16, I-20133 Milano, Italy}

\begin{abstract}
We address characterization of lossy and dephasing channels in the presence of self-Kerr interaction using coherent probes. In particular, we investigate the ultimate bounds to precision in the joint estimation of loss and nonlinearity and of dephasing and nonlinearity. To this aim, we evaluate the quantum Fisher information matrix (QFIM), and compare the symmetric quantum Cramér-Rao bound (QCR) to the bound obtained with Fisher information matrix (FIM) of feasible quantum measurements, i.e., homodyne and double-homodyne detection. 
For lossy Kerr channels, our results show the loss characterization is enhanced in the presence of  Kerr nonlinearity, especially in the relevant limit of small losses and low input energy, whereas the estimation of nonlinearity itself is unavoidably degraded by the presence of loss. In the low energy regime, 
homodyne detection of a suitably optimized quadrature represents a nearly optimal measurement. The Uhlmann curvature does not vanish, therefore loss and nonlinearity can be jointly estimated only with the addition of intrinsic quantum noise. For dephasing Kerr channels, the QFIs of the two parameters are independent of the nonlinearity, and therefore no enhancement is observed. Homodyne and double-homodyne detection are suboptimal for the estimation of dephasing and nearly optimal for nonlinearity. Also in this case, the Uhlmann curvature is nonzero, proving that the parameters cannot be jointly estimated with maximum precision.
\end{abstract}


\maketitle

\section{Introduction}
Nonlinear interactions usually provide an attractive scenario to observe and exploit genuine quantum properties of radiation, as coherence, entanglement and non-Gaussianity \cite{DellAnno2006, Chang2014, Checkova2016, Combes2018, Candeloro2021NonL}.
In this framework, the Kerr effect is a paradigmatic example, being widely studied in quantum optics either at zero \cite{Takatsuji1967, Milburn1986} or finite temperature \cite{Stobinska2008}, as it allows for generation of mesoscopic Schr\"odinger-cat states \cite{Yurke1986, Yurke1988, Miranowicz1990, Paris1999, Jeong2004}.
Furthermore, the presence of Kerr effect has been demonstrated in several physical platforms, ranging from optical media to solid-state systems and circuit quantum electrodynamics \cite{Sizmann1999, Hau1999, Kang2003, Yin2012, Al-Nashi2014, Zhang2017, Zhang2019, Yang2022}.

The Kerr effect is typically observed in optical nonlinear media, such as optical fibers, that exhibit a small but non-negligible third-order susceptibility $\chi^{(3)}$. This makes the refractive index depend on the intensity of the incident light, leading to self-phase modulation, that is the acquirement of a nonlinear intensity-dependent phase shift throughout propagation \cite{Boyd2008}. However, realistic values of Kerr nonlinearity are very small, and decoherence effects, mainly due to photon loss, cannot be neglected, therefore a unitary description of the dynamics is untenable.
As an example, the nonlinearity rate of common fibers ranges from $2 - 5 \times 10^{-7} \, {\rm km}^{-1}$, being several order of magnitudes smaller than the attenuation rate, equal to $0.04 - 0.4 \, {\rm km}^{-1}$ \cite{Kunz2018}. 
In these conditions, the presence of even a small amount of nonlinearity is detrimental for the information capacity of a fiber communication link, operating both in the classical \cite{Mitra2001, Weneger2004, Ellis2010, Essiambre2012, Temprana2015} and quantum regime \cite{Kunz2018}. 
On the contrary, the optical Kerr effect proves itself as a resource for quantum estimation, improving the estimator precision when classical probe states are employed. In particular, it provides enhanced estimation of squeezing and displacement of a Gaussian state \cite{Genoni2009}, increased sensitivity of Michelson interferometry \cite{Luis2015}, improved loss estimation in dissipative bosonic channels \cite{Rossi2016}, and high-precision measurements in atomic systems coupled to an optical cavity \cite{Zidan2019}.

More recently, the presence of Kerr-type effect has been also demonstrated in cavity optomechanical systems, where a single optical cavity mode is coupled to a phononic bath composed of many mechanical oscillator modes \cite{Aspelmeyer2014, Bowen2015, Shang2019, Xu2021, Shang2023}. In this case, the cavity-bath interaction has a twofold consequence on the reduced dynamics of the optical field: the former is phase diffusion, producing decoherence, while the latter is a unitary evolution in the form a Kerr nonlinear self-interaction \cite{Xu2021}.
It is worth to note that, both in optical and optomechanical platforms, self-Kerr interaction arises together with decoherence as the time-evolution of the system proceeds, during either signal propagation throughout the fiber or the roundtrips of the field trapped in cavity. 

This raises the intriguing problem of performing characterization of noisy Kerr channels to assess the impact of nonlinearity in the estimation of the noise parameter and {\em vice versa}.
Given the previous considerations and differently from the approach of Ref.s~\cite{Genoni2009, Rossi2016}, the problem should be recast into the framework of multiparameter quantum metrology \cite{Helstrom1976,Szczykulska2016, Razavian2020, Albarelli2020,Geno08,Brida10}, addressing joint estimation of both noise and nonlinearity to investigate their compatibility, namely whether or not they can be jointly estimated without the introduction of any excess noise of quantum origin.
This task has been only recently carried out in the presence of driven-dissipative Kerr resonators \cite{Asjad2023}, where the coherent driving of the system makes the two parameters asymptotically compatible, whilst the investigation of the relation between noise and nonlinearity in other platforms is still open. 

In this paper, we are going to address the characterization of lossy and dephasing channels in the presence of self-Kerr interaction and using coherent probes.
In particular, we will investigate the ultimate bounds to precision in the joint estimation of loss and nonlinearity and of dephasing and nonlinearity. We evaluate the quantum Fisher information matrix (QFIM), and compare the corresponding symmetric quantum Cramér-Rao bound (QCR) to the bound obtained with Fisher information matrix (FIM) of feasible quantum measurements, i.e., homodyne and double-homodyne detection. 
Our results for lossy Kerr channels show that estimation of loss is enhanced in the presence of  Kerr nonlinearity, especially in the relevant limit of small losses and low input energy, whereas the estimation of nonlinearity itself is unavoidably degraded by the presence of loss. In the low energy regime, 
homodyne detection of a suitably optimized quadrature provides a nearly 
optimal measurement. The Uhlmann curvature does not vanish, showing that 
loss and nonlinearity cannot be jointly estimated without the addition of intinsic quantum noise. For dephasing Kerr channels, the QFIs of the two parameters are independent of the nonlinearity, and therefore no enhancement is observed. Homodyne and double-homodyne detection are suboptimal for the estimation of dephasing and nearly optimal for nonlinearity. Also in this case, the Uhlmann curvature is nonzero, proving that the parameters cannot be jointly estimated without intrinsic quantum noise.

The structure of the paper is the following. In Section~\ref{sec:Metro}, we briefly review multiparameter quantum estimation, introducing the clasical and quantum Cramér-Rao theorem. Thereafter, in Sections~\ref{sec:LossAndKerr} and~\ref{sec:DephAndKerr} we address multiparameter estimation in lossy-Kerr and dephasing-Kerr channels, respectively. In Section~\ref{sec:Resource}, we discuss the physical meaning of our results, whereas Section~\ref{sec:Conc} closes the paper with some concluding remarks.

\section{Basics of multiparameter quantum metrology}\label{sec:Metro}

In a multiparameter metrological problem, the goal is the joint estimation of a set of $N>1$ parameters $\bflambda=\{\lambda_1, \ldots, \lambda_N\}$, being encoded in a quantum state $\rho_{\bflambda}$. The family $\rho_{\bflambda}$ is typically referred to as a quantum statistical model. To infer the values $\{\lambda_\mu\}_\mu$, we perform a quantum measurement described by a positive operator-valued measure (POVM) $\{\Pi_x\}_x$, satisfying $\Pi_x\ge 0$ and $\int d x \, \Pi_x= \hat{\Id}$, $\hat{\Id}$ being the identity operator over the whole Hilbert space. If the measurement is repeated $M$ times, we retrieve a statistical sample of independent and identically distributed outcomes $\textbf{x}=\{x_1, \ldots, x_M\}$, from which we obtain the parameters estimates via an estimator function $\hat{\bflambda}=\hat{\bflambda}(\textbf{x})$ \cite{Helstrom1976,Szczykulska2016, Razavian2020, Albarelli2020}.
Given this scenario, the task is to find the optimal POVM to perform estimation of $\bflambda$ with highest accuracy, namely with the lowest possible uncertainty.

For unbiased estimators, those such that $\mathbb{E}[\hat{\bflambda}]=\bflambda$, the accuracy is quantified by the covariance 
matrix associated with $\hat{\bflambda}$, namely:
\begin{eqnarray}
{\bf V}(\hat{\bflambda}) = \int d{\bf x} \, p(\bf{x}|\bflambda) \left[\hat{\bflambda}({\bf x})- \bflambda \right] \left[\hat{\bflambda}({\bf x})- \bflambda \right]^{\sf T} \, ,
\end{eqnarray}
where $p({\bf x}|\bflambda) = \prod_{j=1}^{M} p(x_j|\bflambda)$ and $p(x|\bflambda) = \Tr [\rho_{\bflambda} \Pi_x]$ is the conditional probability of obtaining outcome $x$ given $\bflambda$. The covariance matrix satisfy the classical Cramér-Rao (CR) bound, defined by the following matrix inequality:
\begin{eqnarray}\label{eq:CCRB}
{\bf V}(\hat{\bflambda}) \ge \left[ M \FI (\bflambda) \right]^{-1} \, ,
\end{eqnarray}
where $\FI (\bflambda)$, with elements 
\begin{eqnarray}\label{eq:FIM}
\F_{\mu \nu}(\bflambda) = \int dx \, p(x|\bflambda) \left[\partial_\mu \log p(x|\bflambda) \right] \left[\partial_\nu \log p(x|\bflambda) \right]\, ,  
\end{eqnarray}
is the classical Fisher information matrix (FIM), $\mu,\nu=1,\ldots, N$, depending on the univariate probability distribution $p(x|\bflambda)$, and $\partial_{\mu(\nu)}$ denotes partial derivatives with respect to $\lambda_{\mu(\nu)}$. 
Moreover, bound~(\ref{eq:CCRB}) may be asymptotically achieved by the maximum likelihood or Bayesian estimators in the limit $M\gg 1$ \cite{Albarelli2020}.

\MN{
Nevertheless, we recall that the FIM depends on the specific POVM $\{\Pi_x\}_x$ being implemented, thereby it is usually considered as a ``classical" quantity. As a consequence, a more general bound can be obtained by optimizing the FIM over all possible quantum measurements, leading to a ``quantum" version of the CR bound, that only depends on the considered statistical model $\rho_{\bflambda}$.
In the single-parameter scenario, this problem was exactly solved by Helstrom, who introduced the quantum Fisher information as the relevant figure of merit, obtained as the maximum Fisher information over all POVMs  \cite{Helstrom1967,Helstrom1976,Szczykulska2016, Razavian2020, Albarelli2020}.
On the contrary, in the multiparameter setting, there exists different possible approaches, corresponding to different figures of merit \cite{Helstrom1967,Yuen1973, Belavkin1976, Holevo1977, Nagaoka1989}. 
}
Here, we consider Helstrom's approach \cite{Helstrom1967} and introduce the symmetric logarithmic derivative (SLD) operators $L_\mu$, $\mu=1,\ldots, N$, defined via the Ljapunov equation 
\begin{eqnarray}\label{eq:Ljap}
\partial_\mu \rho_{\bflambda} = \frac{L_\mu \rho_{\bflambda} + \rho_{\bflambda} L_\mu}{2} \, ,  
\end{eqnarray}
leading to the quantum Fisher information matrix (QFIM) $\QFI(\bflambda)$, with elements \cite{Szczykulska2016, Razavian2020, Albarelli2020, Paris2009}
\begin{eqnarray}\label{eq:QFIM}
\H_{\mu \nu}= \Tr \left[\rho_{\bf \bflambda} \frac{\{L_\mu, L_\nu\}}{2}\right]= \Tr \left[L_\mu \, \partial_\nu \rho_{\bflambda} \right] = \Tr \left[L_\nu \, \partial_\mu \rho_{\bflambda} \right] \, ,
\end{eqnarray}
where $\{A,B\}=AB+BA$ is the anti-commutator of $A$ and $B$. We re-express Equation~(\ref{eq:QFIM}) in a more manageable way by considering the spectral decomposition of $\rho_{\bflambda}$, $\rho_{\bflambda}= \sum_k \rho_k |\phi_k\rangle \langle \phi_k|$ that, combined with~(\ref{eq:Ljap}), leads to \cite{Paris2009}:
 \begin{eqnarray}\label{eq:QFIM1}
\H_{\mu \nu}= 2\sum_{kj} \frac{\langle \phi_k | \partial_\mu \rho_{\bflambda} |\phi_j\rangle \langle \phi_j | \partial_\nu \rho_{\bflambda} |\phi_k\rangle}{\rho_j+\rho_k}\, .
\end{eqnarray}
In particular, in the presence of pure statistical models $ \rho_{\bflambda}= |\psi_{\bflambda}\rangle\langle \psi_{\bflambda}|$, Equation~(\ref{eq:QFIM}) reduces to \cite{Candeloro2021NonL}:
 \begin{eqnarray}\label{eq:QFIpurestate}
\H_{\mu \nu}= 4 \, {\rm Re} \bigg[ \langle \partial_\mu  \psi_{\bflambda}| \partial_\nu  \psi_{\bflambda} \rangle + \langle \partial_\mu  \psi_{\bflambda}|  \psi_{\bflambda} \rangle \langle \partial_\nu  \psi_{\bflambda}|  \psi_{\bflambda} \rangle \bigg]\, .
\end{eqnarray}

The QFIM provides a tighter matrix lower bound on Equation~(\ref{eq:CCRB}), referred to as the SLD-quantum Cramér-Rao (SLD-QCR) bound:
\begin{eqnarray}\label{eq:QCRB}
{\bf V}(\hat{\bflambda})  \ge 
\left[ M \QFI (\bflambda) \right]^{-1} \, .
\end{eqnarray}
Straightforwardly, the former matrix inequalities can be turned into scalar CR bounds by introducing a semipositive definite $N \times N$ weight matrix $\bf W$; then we have: 
$\Tr[{\bf W \, V}] \ge  C_{\FI}({\bf W})$ and $\Tr[{\bf W \, V}]\ge C_{\QFI}({\bf W})$, with $C_{\FI} ({\bf W})= M^{-1} \, \Tr[{\bf W} \, \FI^{-1}]$ and $C_{\QFI} ({\bf W})= M^{-1} \,\Tr[{\bf W}\, \QFI^{-1}]$ \cite{Albarelli2020}.

However, differently from the single-parameter scenario, where the QCR bound  may be achieved by a projective measurement over the SLD eigenstates, in the multiparameter setting, the SLD-QCR bound~(\ref{eq:QCRB}) is not attainable in general, as the SLDs associated with the different parameters may not commute with one another. In this case, the parameters are incompatible, and there is no joint measurement that allows one to estimate all the parameters with the ultimate precision. 
\MN{
Accordingly, one may introduce two other relevant bounds. The former, referred to as the most informative bound, reads $C_{\rm MI}({\bf W})= M^{-1} \min_{\rm POVM} \{\Tr[{\bf W} \, \FI^{-1}]\}$, that, in general, does not coincide with the SLD-QCR bound in the presence of multiple parameters. The latter is the so-called Holevo Cramér-Rao (HCR) bound  $C_{\rm Hol}({\bf W})$, introduced in Ref.~\onlinecite{Holevo1977}, which corresponds to the most informative bound of the asymptotic statistical model, i.e. the minimum FI bound 
achieved by a collective POVM performed on infinitely many copies of the statistical model, namely $\rho_{\bflambda}^{\otimes n}$ with $n\gg 1$ \cite{Albarelli2020, Razavian2020}. In turn, we have $\Tr[{\bf W \, V}] \ge  C_{\FI}({\bf W})  \ge  C_{\rm MI}({\bf W}) \ge  C_{\rm Hol}({\bf W}) \ge C_{\QFI}({\bf W})$; therefore, the HCR bound is usually regarded as the most fundamental scalar bound for multi-parameter quantum estimation.
}

\MN{Given this hierarchy, compatibility of parameters is achieved, at least asymptotically, when the HCR bound saturates the SLD-QCR limit. To this aim, it has been recently proved that \cite{Carollo2018}:
\begin{eqnarray}\label{eq:HolevoHier}
C_{\QFI}({\bf W}) \le C_{\rm Hol}({\bf W}) \le (1+{\cal R}) C_{\QFI}({\bf W}) \, ,
\end{eqnarray}
where the {\em quantumness} parameter ${\cal R}$ 
is given by \cite{Carollo2018,Candeloro2021} 
\begin{eqnarray}\label{eq:R}
{\cal R} = \| i \, \QFI^{-1} \UHL \|_\infty\, ,
\end{eqnarray}
in which $\|{\bf A}\|_\infty$ denotes the largest eigenvalue of the matrix $\bf A$, and
}
$\UHL (\bflambda)$ is the asymptotic incompatibility matrix, also referred to as Uhlmann curvature,
with matrix elements \cite{Carollo2018, Candeloro2021}:
\begin{eqnarray}\label{eq:Uhl}
{\cal U}_{\mu \nu}= -\frac{i}{2} \Tr \left\{\rho_{\bflambda} [L_\mu, L_\nu] \right\} \, ,
\end{eqnarray}
where $[A,B]=AB-BA$ is the commutator of $A$ and $B$, expressed in terms of the eigenstates $\{|\phi_k\rangle\}_k$ of $\rho_{\bflambda}$ as:
\begin{eqnarray}\label{eq:Uhl1}
{\cal U}_{\mu \nu}= 4 \sum_{kj} \frac{\rho_k}{(\rho_k+\rho_j)^2} \, {\rm Im}\bigg[\langle \phi_k | \partial_\mu \rho_{\bflambda} |\phi_j\rangle \langle \phi_j | \partial_\nu \rho_{\bflambda} |\phi_k\rangle \bigg] \, .
\end{eqnarray}
\MN{Equation~(\ref{eq:HolevoHier}) implies that the SLD-QCR bound is saturated iff $\UHL (\bflambda)=0$, referred to as the weak compatibility condition, and the parameters are said to be asymptotically compatible.}
In addition, the ${\cal R}$ quantity satisfies $0\le {\cal R} \le 1$ and ${\cal R}=0$ iff $\UHL (\bflambda)=0$; therefore it provides a measure of asymptotic incompatibility between the parameters. 
In particular, for $N=2$ parameters it reduces to:
\begin{eqnarray}\label{eq:R1}
{\cal R} = \sqrt{\frac{\det \UHL}{\det \QFI}} \qquad \mbox{for $N=2$ parameters.}
\end{eqnarray}
\section{Scenario I: Joint estimation of loss and Kerr nonlinearity}\label{sec:LossAndKerr}
We now address characterization of noisy Kerr channels that provide the evolution of a single-mode optical field, associated with the bosonic operator $a$, $[a,a^\dagger]=1$. In particular, we consider as input a coherent state $|\alpha\rangle= \exp(-|\alpha|^2/2)\sum_{n=0}^\infty \alpha^n/\sqrt{n!} \, |n\rangle$, describing radiation emitted by a laser, where $|n\rangle$ is the Fock state containing $n$ photons, and $\alpha \in \mathbb{C}$ is the field amplitude, such that the mean energy, i.e. mean photon-number, is equal to $\bar{n}=|\alpha|^2$. \cite{Olivares2021}
\MN{We choose coherent states for two reasons: On the one hand, it is an experimentally oriented solution, since many practical realizations of both fiber-optic channels and cavity optomechanical setups exploit coherent radiation, e.g. laser pulses or coherent driving \cite{Asjad2023, Kunz2018}. On the other hand, coherent states represent a benchmark example to perform a comprehensive characterization of noisy nonlinear platforms, that provides a first step towards more advanced analysis, e.g. involving non-classical probes.}

Given the framework reported in the previous Section, we address the joint
estimation of the decoherence parameter and the nonlinearity, which affect the probe state after propagation throughout the channel. 
In more detail, we consider two different scenarios. To begin with, here we consider a lossy-Kerr system, describing propagation of light in optical nonlinear media, e.g. optical fibers \cite{Boyd2008}, referred to as scenario $\caseI$, whereas in the next Section 
we will address a dephasing-Kerr system, referred to as scenario $\caseII$.

For scenario $\caseI$, the time evolution of the system is governed by the following master equation:
\begin{eqnarray}\label{eq:MELoss}
\frac{d \rho}{d t} = - i  \left[\hat{H}_\kappa, \rho \, \right] + \Gamma \, {\cal L}[a] \, \rho \, ,
\end{eqnarray}
where
\begin{eqnarray}\label{eq:HKerr}
\hat{H}_\kappa = \kappa \left(a^\dagger a\right)^2
\end{eqnarray}
is the Hamiltonian describing self-Kerr interaction, $\kappa$ and $\Gamma$ are the Kerr coupling and photon-loss rates, respectively, and ${\cal L}$ is the Lindblad operator such that ${\cal L}[O] \, \rho = O \rho \, O^\dagger - \{O^\dagger O, \rho\}/2$.
Eq. (\ref{eq:MELoss}) may be solved analytically~\cite{Milburn1989,Paris1999,Liu2021} both 
at zero and finite temperature. Given a coherent state $\rho(t=0)=|\alpha\rangle\langle \alpha|$ as input, the quantum state of the system $\rho=\rho(t)$ after time $t$ in its Fock basis expansion is equal to $\rho=\sum_{nm} \rho_{nm} |n\rangle \langle m|$, with:
\begin{eqnarray}\label{eq:rhoI}
\rho_{nm} = \frac{\alpha^n \left(\alpha^{*}\right)^m}{\sqrt{n! m!}} \exp\left\{- \frac{n+m}{2} \, \tau \Delta  - |\alpha|^2 \left[1- \frac{1- e^{-\tau \Delta}}{\Delta}\right] \right\} \, ,
\end{eqnarray}
and
\begin{eqnarray}
\Delta= 1+  \frac{2 i \delta}{\tau} (n-m) \, ,
\end{eqnarray}
where we introduced the quantities:
\begin{eqnarray}\label{eq:ParametersLoss}
\tau= \Gamma t \qquad \mbox{and} \qquad \delta= \kappa t \, ,
\end{eqnarray}
corresponding to the loss and nonlinearity parameter of the channel, respectively.
In particular, given the structure of state~(\ref{eq:rhoI}), we note that the parameters combine themselves in nontrivial way, and the evolution of the system cannot be reported to the dynamics generated separately by $\hat{H}_\kappa$ and $\cal L$. Furthermore, $\tau$ and $\delta$ linearly increase with the interaction time $t$, thus making both the loss and nonlinearity grow for larger lengths of the fiber link. In particular, for long-time interaction, corresponding to the limit $\tau \gg 1$ and $\delta \gg 1$ (with a finite ratio $\delta/\tau$), the decoherence contribution dominates and the system evolves towards the vacuum state, which is the stationary state of the dynamics.

Given these considerations, $\rho$ provides a statistical model with encoded parameters $\blambda=(\tau,\delta)$, whose joint estimation should be investigated. To this aim, in the following we compute the QFIM, that bounds the variance of any estimator, and the Uhlmann curvature, to assess asymptotic compatibility. Thereafter,
we consider few examples of feasible measurements, that is homodyne and double-homodyne detection, and determine their performance by comparing the QFIM and the corresponding FIM.

\subsection{Computation of the QFIM}\label{subsec:QFIMcaseI}

\begin{figure}
\begin{center}
\includegraphics[width=0.48\textwidth]{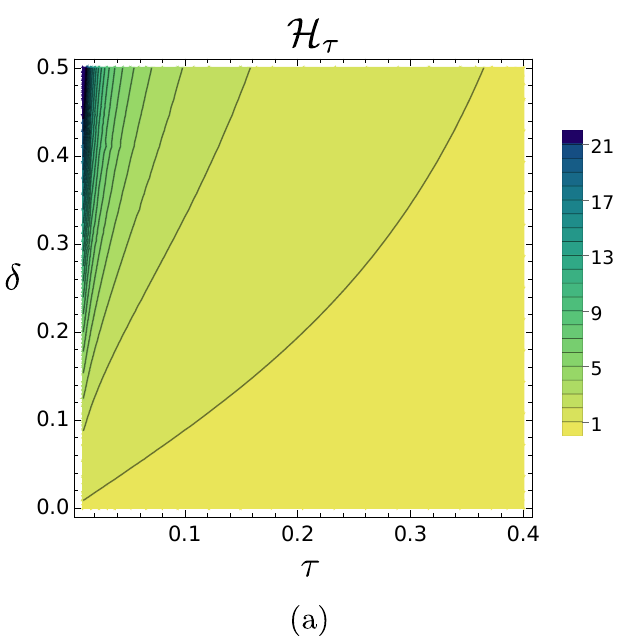} \quad
\includegraphics[width=0.48\textwidth]{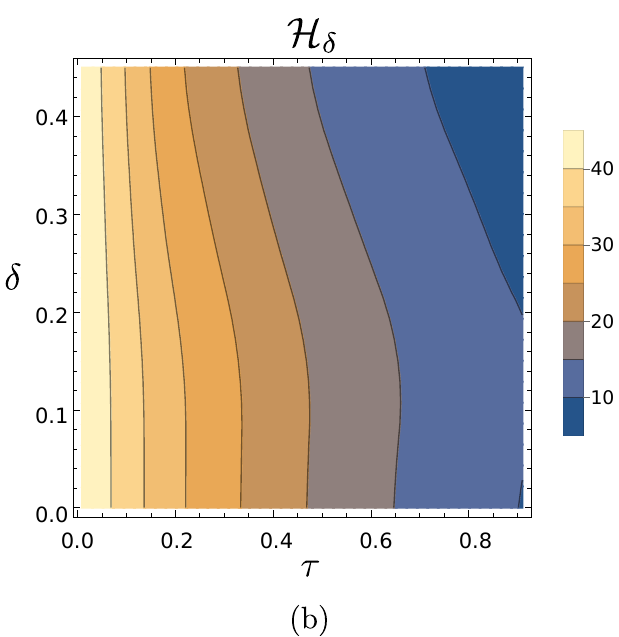}
\end{center}
\caption{Contour plots of the QFIM elements $\H_{\tau}$ (a) and $\H_{\delta}$ (b) for scenario $\caseI$, namely lossy-Kerr systems, as a function of the loss parameter $\tau=\Gamma t$ and the nonlinearity parameter $\delta= \kappa t$, for an input coherent state with amplitude $\alpha=1$. The presence of a nonzero nonlinearity enhances the loss estimation, the effect being accentuated for small values of $\tau$, whereas the presence of loss reduces $\H_{\delta}$ monotonically for all values of $\delta$.}\label{fig:01-QFILoss}
\end{figure}

We start by computing the QFIM $\QFI$. To this aim, we consider the operator $\partial_\mu \rho=\sum_{nm}( \partial_\mu \rho_{nm} )|n\rangle \langle m| $, computed by deriving~(\ref{eq:rhoI}) with respect to parameter $\mu=\tau,\delta$, and retrieve the QFIM elements from Equation~(\ref{eq:QFIM1}), where the eigenvalues $\{\rho_k\}_k$ and eigenstates $\{|\phi_k\rangle \}_k$ of the statistical model $\rho$ are obtained via numerical diagonalization.
\MN{To perform the calculation, we truncated the Hilbert space to a suitable dimension $d$ such that $1-\Tr[\rho]<10^{-5}$. Subsequent numerical checks show that, with the chosen parameters, our results are accurate up to the fifth significant digit, corresponding to a relative error about $10^{-5}$, which provides a sufficient precision level for the purposes of this study.}

Plots of the diagonal elements $\H_{\tau}= \H_{\tau \tau}$ and $\H_{\delta}= \H_{\delta \delta}$, corresponding to the maximum precision associated with single-parameter estimation of $\tau$ and $\delta$, are reported in Figure~\ref{fig:01-QFILoss}(a) and (b), respectively, as a function of $\tau$ and $\delta$ for fixed input coherent state amplitude $\alpha=1$. The behaviour is qualitatively similar for all $\alpha$.
As we can see, the presence of losses is detrimental for the nonlinearity estimation, as $\H_{\delta}$ monotonically decreases with $\tau$ for all values of $\delta$. In fact, increasing $\tau$ suppresses the matrix elements $\rho_{nm}$ with $n,m>0$, thus progressively reducing the sensitivity of state~(\ref{eq:rhoI}) to $\delta$.
On the contrary, a nontrivial effect emerges in the estimation of the loss parameter. In fact, as already showed in \cite{Rossi2016}, the presence of a nonzero Kerr susceptibility makes $\H_{\tau}$ increase, enhancing the sensitivity of $\rho$ to parameter $\tau$ and proving the Kerr effect as a beneficial tool. The enhancement is more accentuated in the limit of small $\tau$, corresponding to low-loss optical fibers or short link lengths. The physical interpretation of this result will be discussed thereafter in Section~\ref{sec:Resource}. 
Interestingly, analytic results can be retrieved in the limits $\tau \ll 1$ and $\delta \ll 1$, in which the term $1-e^{-\tau \Delta}$ in Equation~(\ref{eq:rhoI}) can be expanded up to the second order, i.e. $1-e^{-\tau \Delta}\approx \tau \Delta -(\tau \Delta)^2/2$, and the matrix elements of $\rho$ factorize as $\rho_{nm} = c_n c_m^*$, with:
\begin{eqnarray}
c_{n} = \frac{\alpha^n}{\sqrt{n!}} \exp\left\{- \frac{\tau n}{2}  - i \delta n^2 - |\alpha|^2 \left[\frac{e^{-\tau}}{2} + i \tau \delta \, n \right] \right\} \, .
\end{eqnarray}
That is, when $\tau \ll 1$ and $\delta \ll 1$, the state of the system remains approximately pure, $\rho=|\psi\rangle \langle \psi|$, with $|\psi\rangle = \sum_n c_n |n\rangle$. In turn, Equation~(\ref{eq:QFIpurestate}) holds and we have:
\begin{subequations}
\begin{eqnarray}
\H_{\tau} &\approx& \H_\tau^{(0)} \left(1+ 4  |\alpha|^4 \delta^2\right) \, , \\[1ex]
\H_{\delta} &\approx& \H_\delta^{(0)} -4  |\alpha|^2\left(1+10  |\alpha|^2+8  |\alpha|^6\right) \tau \, ,
\end{eqnarray}
\end{subequations}
where $\H_\tau^{(0)}= e^{-\tau} |\alpha|^2$ is the QFI for the single-parameter estimation of $\tau$ in the absence of nonlinearity, when $\rho$ becomes equal to the rescaled coherent state $|e^{-\tau/2}\alpha\rangle$, whereas $\H_\delta^{(0)}= 4 |\alpha|^2 (1+6 |\alpha|^2+4 |\alpha|^4)$ is the QFI for $\delta$ in the absence of loss, in which case the encoded state is equal to $\exp(-i \delta (a^\dagger a)^2)|\alpha\rangle$ \cite{Monras2007, Adesso2009, Cheng2014}. In summary, in the pure-state approximation limit, the nonlinearity induces a quadratic enhancement in the loss estimation, whereas the loss produces a linear reduction of the nonlinearity QFI, proving $\rho$ to be fragile with respect to nonlinearity in the presence of decoherence.
Finally, in accordance with the previous considerations, we also compute the correlation QFIM term $\H_{\tau \delta}$, which decreases with $\tau$ if $\delta$ is fixed, while increases with $\delta$ for fixed $\tau$.

\begin{figure}
\begin{center}
\includegraphics[width=0.48\textwidth]{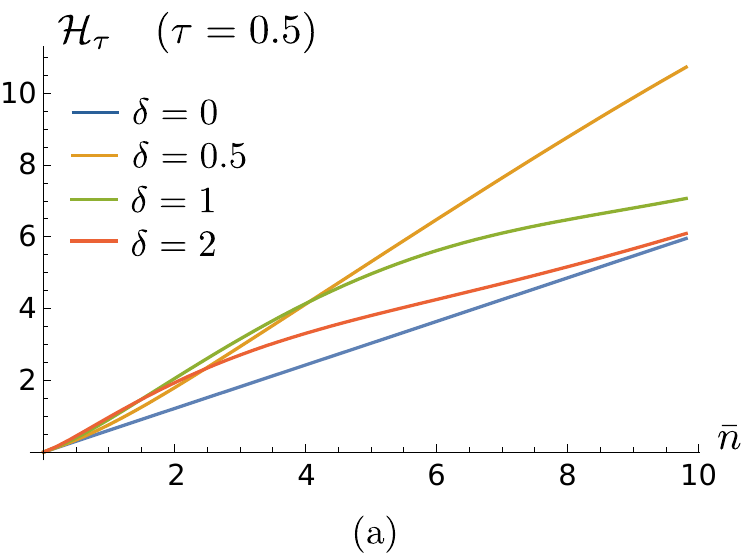} \quad
\includegraphics[width=0.48\textwidth]{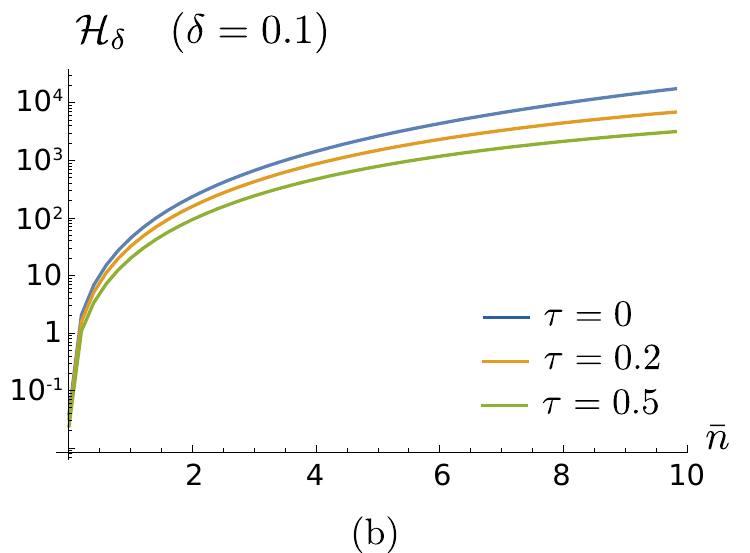}
\end{center}
\caption{(a) Plot of the loss-QFI $\H_{\tau}$ for scenario $\caseI$, namely lossy-Kerr systems, as a function of the input coherent state energy $\bar{n}=|\alpha|^2$ for loss parameter $\tau=0.5$ and different values of nonlinearity $\delta$. (b) Log plot of the nonlinearity-QFI $\H_{\delta}$ for scenario $\caseI$ as a function of $\bar{n}$ for $\delta=0.1$ and different $\tau$.}\label{fig:02-QFILoss-vs-alpha}
\end{figure}

\begin{figure}
\begin{center}
\includegraphics[width=0.48\textwidth]{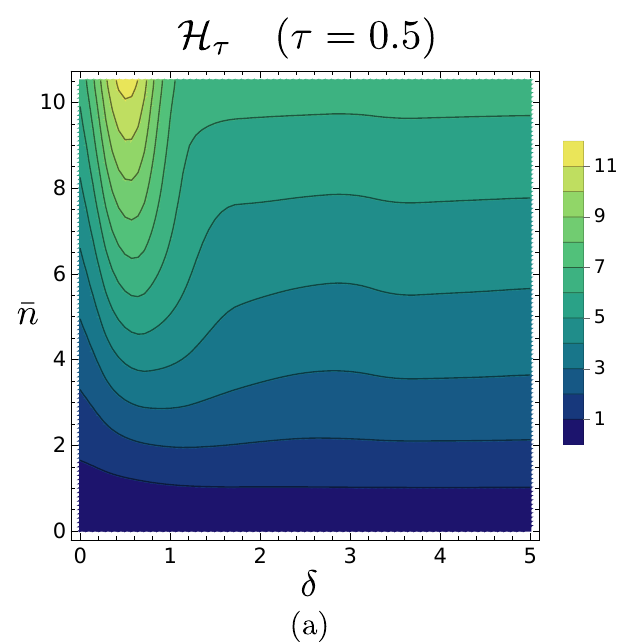} \quad
\includegraphics[width=0.48\textwidth]{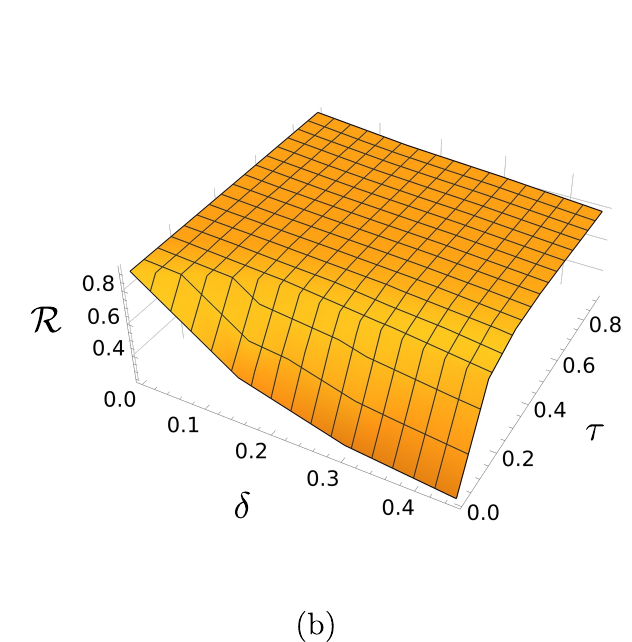}
\end{center}
\caption{(a) Contour plot of the loss-QFI $\H_{\tau}$ for scenario $\caseI$, namely lossy-Kerr systems, as a function of the nonlinearity $\delta$ and the input energy $\bar{n}$ for $\tau=0.5$. For each $\bar{n}$ there exists a finite value $\delta_{\rm max} <\infty$ that maximizes the loss-QFI. (b) Plot of the quantumness ${\cal  R}$ for scenario $\caseI$ as a function of the channel parameters $\tau$ and $\delta$ for an input energy $\bar{n}=1$. ${\cal  R}$ is always nonzero and saturates in the limit of large $\tau$, proving the two parameters to be incompatible.}\label{fig:03-QFILossContour}
\end{figure}

Furthermore, the enhancement in the loss estimation, being about a factor $20$ for $\alpha=1$, is progressively reduced as the energy of the input probe state $\bar{n}=|\alpha|^2$ increases; whereas the reduction of the nonlinearity QFI $\H_\delta$ is more accentuated for higher $\bar{n}$. This makes it worth to investigate the QFIM dependence on the input coherent state energy. In particular, given~(\ref{eq:rhoI}), we may safely assume that $\alpha>0$, as the presence of a complex coherent amplitude only adds a phase shift to the matrix elements $\rho_{nm}$, being insensitive to both the loss and Kerr parameters.
Figure~\ref{fig:02-QFILoss-vs-alpha}(a) and (b) reports $\H_{\tau}$ and $\H_{\delta}$ as a function of $\bar{n}$, respectively.
We observe that the sensitivity to $\delta$ increases monotonically with $\bar{n}$ for all $\tau$. If $\tau=0$, we have $\H_{\delta}=\H_\delta^{(0)}= 4 \bar{n} (1+6 \bar{n}+4 \bar{n}^2)$; when $\tau>0$, $\H_{\delta}$ decreases, the reduction being more accentuated in the high-energy regime, due to the suppression term proportional to $|\alpha|^2$ in~(\ref{eq:rhoI}).
Instead, $\H_{\tau}$ shows a non-monotonic behaviour when $\delta>0$. For $\delta=0$, we retrieve the single-parameter scenario $\H_\tau=\H_\tau^{(0)}= e^{-\tau} \, \bar{n}$, that is shot-noise scaling linearly increasing with the energy, whilst, in the presence of Kerr nonlinearity, the QFI $\H_\tau$ is increased in the low-energy regime until to reach a maximum, after which it decreases and re-approaches $\H_\tau^{(0)}$ in the asymptotic limit $\bar{n}\gg 1$. Conversely, for each $\bar{n}$, there exists a finite value of nonlinearity $\delta_{\rm max} <\infty$ that maximizes $\H_\tau$, whilst increasing further $\delta$ turns out to be useless, as enlightened in Figure~\ref{fig:03-QFILossContour}(a), that reports the loss-QFI as a function of $\delta$ and $\bar{n}$ for $\tau=0.5$, corresponding to \MN{almost} $10 \, {\rm km}$ transmission in common fibers \cite{Banaszek2020}.

Finally, we assess the compatibility of the two parameters by computing the Uhlmann curvature $\UHL$, equal to:
\begin{eqnarray}\label{eq:UhlmannLoss}
\UHL =
\left(
\begin{array}{cc} 
0 & {\cal U}_{\tau \delta} \\
- {\cal U}_{\tau \delta} & 0 \\
\end{array}
\right) \, ,
\end{eqnarray}
where the off-diagonal term ${\cal U}_{\tau \delta}$ is computed from Equation~(\ref{eq:Uhl1}) with the analogous method adopted for the QFIM. 
The numerical results show that ${\cal U}_{\tau \delta}$ is a decreasing function of $\tau$, being always nonzero for all $\tau$ and $\delta$. In turn, the two parameters cannot be jointly estimated with maximum precision. To quantify their incompatibility, we consider the quantumness ${\cal  R}=\sqrt{\det \UHL / \det \QFI}$, depicted in Figure~\ref{fig:03-QFILossContour}(b). We have ${\cal  R}> 0$ and, in the limit of large $\tau$, ${\cal  R}$ exhibits a weak dependence on $\delta$ and saturates. The saturation value is $ {\cal  R}_\infty\approx 0.8$ for $\bar{n}=1$ and increases for higher $\bar{n}$, suggesting the Holevo scalar bound introduced in Section~\ref{sec:Metro} to be not close to the SLD-QFI bound.

\subsection{Performance of feasible measurements}\label{subsec:FIMcaseI}
As we demonstrated above, the channel parameters $\tau$ and $\delta$ cannot be jointly estimated with maximum precision, thus the SLD-QCR bound is not attainable. Then, it becomes interesting to consider some feasible measurement schemes, that could be easily implemented in practice, and compare their performance with the ultimate bounds provided by the QFIM. Within this class, Gaussian measurements, such as homodyne and double-homodyne detection, provide the typical example of feasible detection strategies for optical signals \cite{Serafini2017}. In principle, direct detection, i.e. photon-number measurement, could be also considered. Nevertheless, the photon-number distribution associated with state $\rho$ is retrieved from its diagonal matrix elements, namely $p(n|\bflambda)=\langle n | \rho|n\rangle= \rho_{nn}$, being independent of the nonlinearity $\delta$, therefore the joint estimation problem is trivially reported to the single-parameter estimation of $\tau$.

\begin{figure}
\begin{center}
\includegraphics[width=0.48\textwidth]{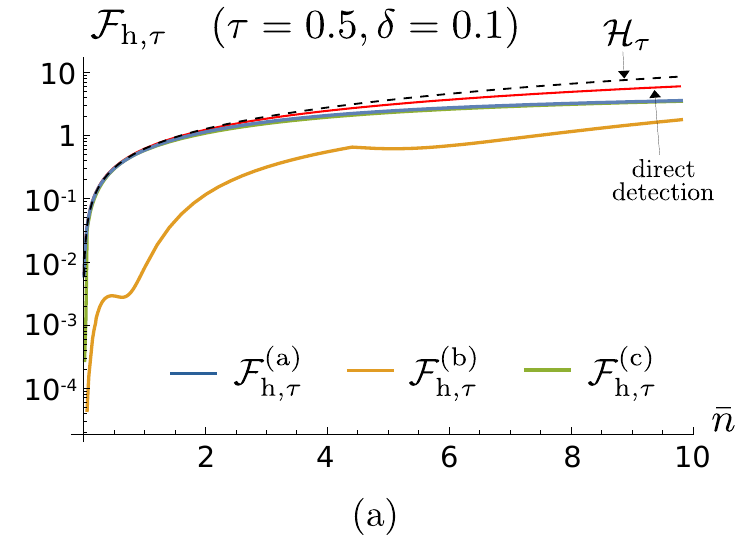} \quad
\includegraphics[width=0.48\textwidth]{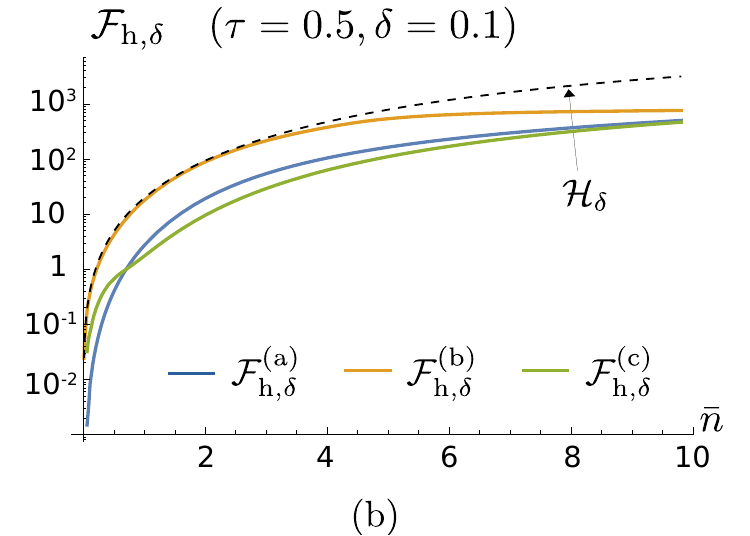}
\end{center}
\caption{Log plot of the loss-FI $\F^{(\p)}_{\HD,\tau}$ for homodyne detection under scenario $\caseI$, namely lossy-Kerr systems, (a) and the nonlinearity-FI $\F^{(\p)}_{\HD,\delta}$ (b), $\rm p=a,b,c$, as a function of $\bar{n}$. 
Homodyne detection proves itself to be nearly optimal for both loss and nonlinearity estimation in the low-energy regime. \MN{The red line in panel (a) is the loss-FI for direct detection, equal to $\H_\tau^{(0)}=e^{-\tau} \bar{n}$.} We set the values $\tau=0.5$ and $\delta=0.1$ for the loss and nonlinearity parameters, respectively.}\label{fig:04-HD-FILoss-vs-alpha}
\end{figure}

\begin{figure}
\begin{center}
\includegraphics[width=0.55\textwidth]{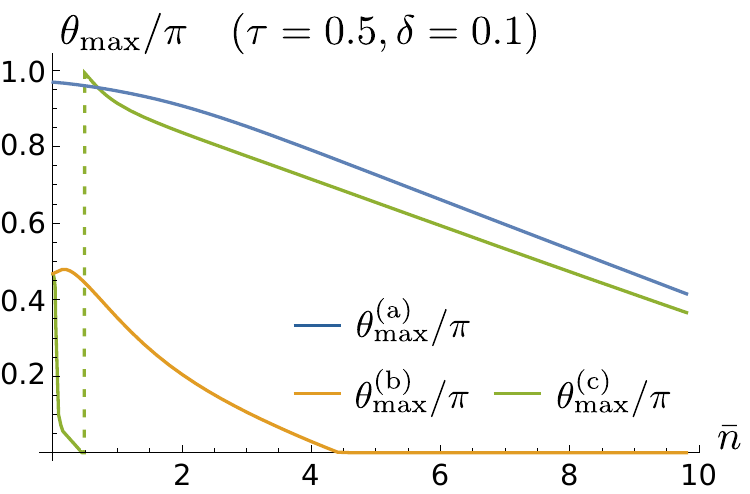}
\end{center}
\caption{Plot of the optimized quadrature phases $\theta_{\rm max}^{(\p)}/\pi$, $\rm p=a,b,c$, for homodyne detection under scenario $\caseI$, namely lossy-Kerr systems, as a function of $\bar{n}$. We set the values $\tau=0.5$ and $\delta=0.1$ for the loss and nonlinearity parameters, respectively.}\label{fig:05-PhiMAXLoss}
\end{figure}

To begin with, we compute the FIM $\FI_\HD=(\F_{\HD,\mu\nu})_{\mu,\nu}$ associated with homodyne detection, corresponding to the measurement of the field quadrature:
\begin{eqnarray}
\hat{x}_\theta= \sigma_0 \left( a e^{-i\theta} + a^\dagger e^{i\theta} \right) \, ,
\end{eqnarray}
where $0\le \theta<\pi$ determines the phase of the probed quadrature, and $\sigma_0^2$ is the shot-noise variance, corresponding to vacuum fluctuations, such that $\langle 0 |\hat{x}_\theta^2|0\rangle = \sigma_0^2$\cite{Olivares2021}. We also remind that homodyne detection of $\hat{q}=\hat{x}_0=\sigma_0 (a+a^\dagger)$ provides the optimal measurement for the loss estimation in the absence of nonlinearity, saturating the single-parameter SLD-QCR bound \cite{Rossi2016}, while it is only suboptimal for the nonlinearity estimation, due to the non-Gaussian nature of the Kerr effect.
Performing detection of $\hat{x}_\theta$ is equivalent to the $1$-rank projective measurement $\Pi_x(\theta)= |x\rangle_\theta \langle x|$, where:
\begin{eqnarray}
|x\rangle_\theta = \frac{e^{-x^2/2}}{\pi^{1/4}} \sum_{n=0}^{\infty} \frac{H_n(x)}{\sqrt{2^n n!}} \, e^{-i n\theta} \, |n\rangle \, ,
\end{eqnarray}
$H_n(x)$ being the $n$-th Hermite polynomial \cite{Olivares2021}. In turn, the homodyne probability distribution reads 
\begin{eqnarray}\label{eq:HomodyneProb}
p_\theta(x|\bflambda) &=& {}_\theta\langle x|\rho |x\rangle_\theta \nonumber \\[1ex]
&=& \frac{e^{-x^2}}{\sqrt{\pi}} \sum_{nm} \rho_{nm} \, \frac{H_n(x) H_m(x)}{\sqrt{2^{n+m} n!m!}} \, e^{i (n-m)\theta} \,,
\end{eqnarray}
and the corresponding FIM is numerically retrieved via Equation~(\ref{eq:FIM}).
In particular, we optimize the quadrature phase $\theta$ to achieve the maximum performance. Due to the incompatibility of $\tau$ and $\delta$, we identify three different cases:
\begin{itemize}
\item case (a): we optimize $\theta$ to maximize precision on the loss estimation, i.e. maximizing the loss-FI $\F_{\HD,\tau}= \F_{\HD,\tau\tau}$;
\item case (b): we optimize $\theta$ to maximize precision on the nonlinearity estimation, i.e. maximizing the nonlinearity-FI $\F_{\HD,\delta}= \F_{\HD,\delta\delta}$;
\item case (c): we optimize $\theta$ to maximize precision on the sum of the mean square errors for each parameter, i.e. maximizing ${\cal C}_\HD= 1/\Tr[\FI_{\HD}^{-1}]$, corresponding to minimize the inverse of trace of the inverse FIM.
\end{itemize}
Plots of $\F^{(\p)}_{\HD,\tau}$ and $\F^{(\p)}_{\HD,\delta}$, $\rm p= a,b,c$, as a function of the input energy $\bar{n}$ are depicted in Figure~\ref{fig:04-HD-FILoss-vs-alpha}(a) and (b), respectively, compared to the corresponding QFIs. In both the cases, we see that, if properly optimized, the homodyne FI is close to the corresponding QFI for sufficiently small energy $\bar{n}$, while the separation increases for higher $\bar{n}$. With the chosen values of $\tau$ and $\delta$ we have $\F^{(\a)}_{\HD,\tau} \gtrsim 0.9 \, \H_{\tau}$ for $\bar{n}\lesssim2$, whereas $\F^{(\b)}_{\HD,\delta} \gtrsim 0.8 \,\H_{\tau}$ for $\bar{n}\lesssim 4$. Moreover, we note that $\F^{(\c)}_{\HD,\tau} \approx \F^{(\a)}_{\HD,\tau}$, while the three subcases lead to distinct FIs for the nonlinearity estimation.
The plots of the optimized phase $\theta_{\rm max}^{(\p)}$, $\rm p=a,b,c$, are depicted in Figure~\ref{fig:05-PhiMAXLoss}, where the jump that appears is only consequence of the $\pi$-periodicity of the phase, which, by construction, has been constrained in the interval $0\le \theta <\pi$. Given this consideration, both $\theta_{\rm max}^{(\rm a)}$ and $\theta_{\rm max}^{( \rm c)}$ are decreasing function of $\bar{n}$. 
On the contrary, in case (b), the optimized phase is non-monotonic: it is $\theta_{\rm max}^{(\b)} \approx \pi/2$ for $\bar{n}\ll 1$, while it drops to $0$ for $\bar{n}\ge \bar{n}_0$, showing quadrature $\hat{q}$ to be the best one for the nonlinearity estimation. The discontinuity in the derivative of $\theta_{\rm max}^{(\b)}$ at $\bar{n}_0$ is then reflected in the cusp of $\F^{(\b)}_{\HD,\tau}$, see Figure~\ref{fig:04-HD-FILoss-vs-alpha}(a).
\MN{Interestingly, our numerical results also show that homodyning $\hat{q}$, which provides the optimal measurement for the loss estimation in the absence of nonlinearity, becomes strongly suboptimal when $\delta>0$. This is a consequence of the phase-sensitive behaviour of the Kerr interaction, which induces distortions of coherent states in the phase space by introducing both nonlinear phase noise and non-Gaussian squeezing along a proper direction \cite{Stobinska2008,Kunz2018}, thus making quadrature $\hat{q}$ less sensitive to parameter $\tau$. Given these considerations, one may wonder which is the performance of direct detection, being optimal too when $\delta=0$, compared to the optimized homodyne cases (a), (b) and (c). As briefly mentioned before, direct detection is insensitive to the nonlinearity, therefore its loss-FI is identical to that of the pure-loss scenario, namely equal to $\H_\tau^{(0)}=e^{-\tau}\bar{n}$. Even though its optimality does not hold when $\delta>0$, photon-number measurement outperforms homodyne detection in all cases and all energy regimes, as we can see from Figure~\ref{fig:04-HD-FILoss-vs-alpha}(a), reducing the separation with respect to $\H_\tau$.
However, because of the insensitivity to $\delta$, this kind of measurement yields useful application only for the single-parameter loss estimation, whereas its adoption in the multiparameter setting turns out to be useless.
}

\begin{figure}
\begin{center}
\includegraphics[width=0.48\textwidth]{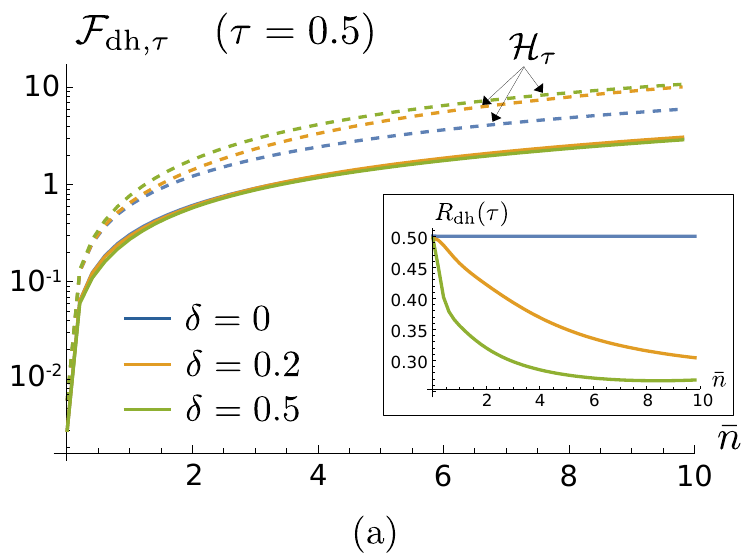} \quad
\includegraphics[width=0.48\textwidth]{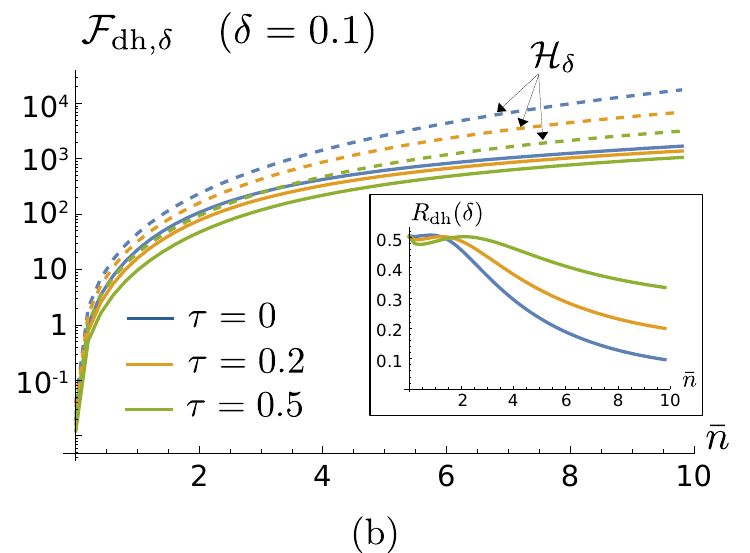}
\end{center}
\caption{(a) Log plot of the loss-FI $\F_{\HET,\tau}$ for DH detection under scenario $\caseI$, namely lossy-Kerr systems, and plot of the relative ratio $R_\HET(\tau)= \F_{\HET,\tau}/ \H_{\tau}$ (in the inset) as a function of $\bar{n}$ for $\tau=0.5$ and different values of the nonlinearity $\delta$. (b) Log plot of the nonlinearity-FI $\F_{\HET,\delta}$ and plot of the relative ratio $R_\HET(\delta)= \F_{\HET,\delta}/ \H_{\delta}$ (in the inset) as a function of $\bar{n}$ for $\delta=0.1$ and different values of the loss parameter $\tau$.}\label{fig:06-HET-FILoss-vs-alpha}
\end{figure}

Now, we address the second example of feasible POVM, that is double-homodyne (DH) detection. It corresponds to joint measurement of the two orthogonal quadratures, obtained by splitting the incoming signal in two copies at a balanced beam splitter and then homodyning $\hat{q}=\hat{x}_0$ and $\hat{p}=\hat{x}_{\pi/2}$ on the transmitted and reflected branch, retrieving a pair of real outcomes $(x,y)\in\mathbb{R}^2$, respectively. The main consequence of the signal splitting is the introduction of a ineludible excess noise on both the output quadrature statistics, equal to $\sigma_0^2$, that guarantees joint measurement without violation of the Heisenberg's uncertainty principle \cite{Banaszek2020}. This excess noise makes DH only suboptimal for the loss estimation in the absence of nonlinearity, its associated FI being exactly one half of the QFI achieved by single-homodyne of $\hat{q}$ \cite{Rossi2016}.
Equivalently, DH detection is described as a $1$-rank (non-orthogonal) projection on coherent states, with associated POVM $\Pi_{x,y}= |\zeta_{x,y}\rangle \langle \zeta_{x,y}|/2\pi$, where $|\zeta_{x,y}\rangle$ is a coherent state with amplitude $\zeta_{x,y}=(x+iy)/\sqrt{2}$, such that $\int dxdy \, \Pi_{x,y}= \hat{\Id}$.
In turn, the corresponding probability distribution given state $\rho$ reads:
\begin{eqnarray}\label{eq:DoubleHomodyneProb}
p_\HET(x,y|\bflambda) 
&=& \frac{1}{2\pi}\langle \zeta_{x,y} |\rho | \zeta_{x,y}\rangle \nonumber \\[1ex]
&= &\frac{e^{-(x^2+y^2)/2}}{2\pi} \sum_{nm}  \frac{\rho_{nm}}{\sqrt{n!m!}} \left(\frac{x-i y}{\sqrt{2}}\right)^n \left(\frac{x+i y}{\sqrt{2}}\right)^m \,,
\end{eqnarray}
from which we compute the FIM $\FI_\HET=(\F_{\HET,\mu\nu})_{\mu,\nu}$ thanks to~(\ref{eq:FIM}).
Figure~\ref{fig:06-HET-FILoss-vs-alpha}(a) and (b) shows the resulting FIM elements $\F_{\HET,\tau}$ and $\F_{\HET,\delta}$, respectively, as a function of the input energy $\bar{n}$. As we can see, DH detection is weakly dependent on the nonlinearity $\delta$, as the value of $\F_{\HET,\tau}$ is almost the same for different $\delta$, whereas it proves itself quite robust against the loss, since $\F_{\HET,\delta}$ decreases more slowly than $\H_{\delta}$ when $\tau$ is increased. To better enlighten these effects, we consider the relative ratios:
\begin{eqnarray}
R_\HET(\mu) = \frac{\F_{\HET,\mu}}{\H_{\mu}} \, , \quad \mu=\tau,\delta \, ,
\end{eqnarray}
depicted in the insets of Figure~\ref{fig:06-HET-FILoss-vs-alpha}. As regards the loss estimation, if $\delta=0$, we have $R_\HET(\tau)=0.5$, retrieving the known result for single-parameter estimation, whereas for $\delta>0$ the ratio decreases, since the loss-QFI is enhanced while the FI value $\F_{\HET,\tau}$ remains almost stable. On the contrary, in estimating the Kerr nonlinearity we identify two different regimes. If $\bar{n}\ll 1$, $R_\HET(\delta)$ decreases with $\tau$, while, on the other hand, for high enough energy the situation is reversed and $R_\HET(\delta)$ gets higher values for larger losses, showing higher robustness against decoherence. Beside this, DH is always suboptimal, and the ratio $R_\HET(\mu)$, $\tau,\delta$ is lower than $0.5$ in all conditions.

\begin{figure}
\begin{center}
\includegraphics[width=0.58\textwidth]{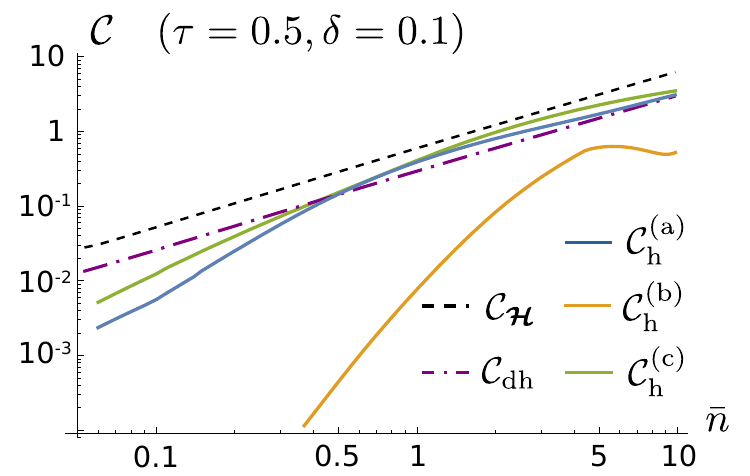}
\end{center}
\caption{Log-log plot of the trace scalar bounds for scenario I, namely lossy-Kerr systems, that is the SLD-QFIM bound ${\cal C}_\QFI= 1/\Tr[\QFI^{-1}]$, and the FIM bounds for homodyne detection, ${\cal C}^{(\p)}_\HD= 1/\Tr[(\FI_\HD^{(\p)})^{-1}]$, $\rm p=a,b,c$, and DH detection, ${\cal C}_\HET=1/\Tr[\FI_\HET^{-1}]$, as a function of $\bar{n}$. We set the values $\tau=0.5$ and $\delta=0.1$ for the loss and nonlinearity parameters, respectively.}\label{fig:07-TraceLoss}
\end{figure}

To conclude, we also compare the performance of the previous measurement schemes in terms of scalar CR bounds. In particular, we choose the weight matrix ${\bf W}=\Id_2$, $\Id_2$ being the $2\times 2$ identity matrix, in which case the figure of merit is the trace of the inverse FIMs. Then, the covariance matrix $\bf V$ of an unbiased estimator satisfies $\Tr[{\bf V}] \ge 1/{\cal C} \ge 1/{\cal C}_\QFI$, with ${\cal C}^{(\p)}_{\HD}=1/\Tr[(\FI^{(\p)}_\HD)^{-1}]$, $\rm p=a,b,c$, for homodyne detection, ${\cal C}_{\HET}= 1/\Tr[\FI_\HET^{-1}]$ for DH detection, and ${\cal C}_\QFI= 1/\Tr[\QFI^{-1}]$, in which, for the sake of simplicity, we set the number of measurements equal to $M=1$.
Plots of the trace scalar bounds are reported in Figure~\ref{fig:07-TraceLoss}.
Consistently with the former results, none of the considered measurements is able to reach ${\cal C}_\QFI$. Nevertheless, as we can see, homodyne detection for case (c), is nearly optimal for a sufficiently high input energy, being beaten by DH detection only in the low-energy limit $\bar{n}\ll 1$, when ${\cal C}_{\HET} \ge {\cal C}^{(\c)}_{\HD}$. Remarkably, the performance of case (a) is close to (c), whilst case (b) is strongly suboptimal. This is mainly due to the different sensitivity to $\tau$ and $\delta$ of the off diagonal element of matrix $\FI_{\HD}$. Finally, in the limit of high $\bar{n}$, DH detection approaches case (a)-homodyne detection, as ${\cal C}_{\HET} \approx {\cal C}^{(\a)}_{\HD}$, proving itself as a versatile solution, that does not involve optimization of the experimental setup, in all energy regimes, regardless its suboptimality.

\section{Scenario II: Joint estimation of dephasing and Kerr nonlinearity}\label{sec:DephAndKerr}
We now move on to the second scenario under investigation, and perform characterization of noisy systems in the presence of Kerr nonlinearity and dephasing produced by phase diffusion \cite{Genoni2011,Notarnicola2022}. This is a typical situation emerging in quantum optomechanical systems, consisting in an optical cavity whose mirrors experience quantized vibrations at acoustic frequencies. The mechanical oscillations of the mirrors, then, induce a change in the effective length of the cavity and, accordingly, in its proper frequency, resulting in an overall interaction between the corresponding optical and mechanical bosonic modes \cite{Bowen2015}. In particular, if the acoustic modes are excited in a thermal state, the reduced dynamics of the optical cavity field is equivalent to a phase diffusion master equation with an effective self-Kerr unitary interaction, where both the noise and nonlinearity parameter depend on the strength of the optomechanical interaction \cite{Xu2021}.
That is, the evolution of the optical state $\rho$ of the cavity field follows:
\begin{eqnarray}\label{eq:MEDeph}
\frac{d \rho}{d t} = - i  \left[\hat{H}_\kappa, \rho \, \right] + \gamma \, {\cal L}[a^\dagger a] \, \rho \, ,
\end{eqnarray}
with the $\hat{H}_\kappa$ in Equation~(\ref{eq:HKerr}) and $\kappa$ and $\gamma$ being the effective Kerr coupling and phase diffusion rates, respectively.
The solution to the master equation at time $t$, $\rho=\rho(t)$, is straightforwardly obtained by expanding $\rho$ on the Fock basis, namely $\rho= \sum_{nm} \rho_{nm} |n\rangle \langle m|$, leading to:
\begin{eqnarray}\label{eq:rhoII}
\rho_{nm} = \frac{\alpha^n \left(\alpha^{*}\right)^m}{\sqrt{n! m!}} \exp\left\{- |\alpha|^2 - i \delta (n^2-m^2) - \frac{\sigma^2}{2} (n-m)^2 \right\} \, ,
\end{eqnarray}
with the coherent state $|\alpha\rangle$ as input, and where:
\begin{eqnarray}\label{eq:ParametersLoss}
\sigma= \sqrt{\gamma t} \qquad \mbox{and} \qquad \delta= \kappa t \, ,
\end{eqnarray}
represent the dephasing parameter, also referred to as phase noise amplitude, and the nonlinearity parameter, respectively.
Differently from scenario $\caseI$, both the Hamiltonian and Lindblad dynamics are generated by the photon-number operator, thus solution~(\ref{eq:rhoII}) can be re-expressed as:
\begin{eqnarray}\label{ew:rhoCOV}
\rho= e^{-i \delta (a^\dagger a)^2} \, \varrho_\sigma \, e^{-i \delta (a^\dagger a)^2} \, ,
\end{eqnarray}
$\varrho_\sigma=e^{-|\alpha|^2} \sum_{nm} e^{- \sigma^2 (n-m)^2/2} \, \alpha^n (\alpha^{*})^m/\sqrt{n! m!}$ being a dephased coherent state, namely as the subsequent application of a dephasing completely-positive map followed a unitary Kerr evolution.
In turn, now, no pure-state approximation can be carried out, since the presence of phase diffusion makes $\rho$ mixed for all $\sigma>0$.
The stationary state of the dynamics, achieved in the limits $\sigma \gg 1$ and $\delta \gg 1$, is the phase-averaged (PHAV) state \cite{Allevi2012}:
\begin{eqnarray}
\rho_{\rm PHAV}= e^{-|\alpha|^2} \sum_n \frac{|\alpha|^{2n}}{n!} \, |n\rangle \langle n|\, ,
\end{eqnarray}
corresponding to a Poisson-distributed ensemble of Fock states, being insensitive to the nonlinearity.
In this scenario, we have a statistical model $\rho$ encoding parameters $\blambda=(\sigma,\delta)$. As before, we compute the QFIM and the Uhlmann curvature, to assess the ultimate precision limits. Then, we further compute the FIM associated with homodyne and DH detection, comparing their performances to the SLD-QCR bound.

\subsection{Computation of the QFIM}\label{subsec:QFIMcaseII}

\begin{figure}
\begin{center}
\includegraphics[width=0.48\textwidth]{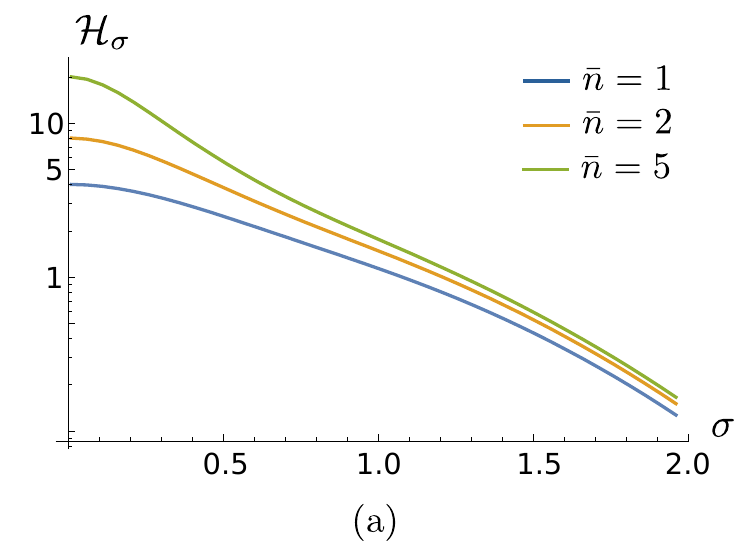} \quad
\includegraphics[width=0.48\textwidth]{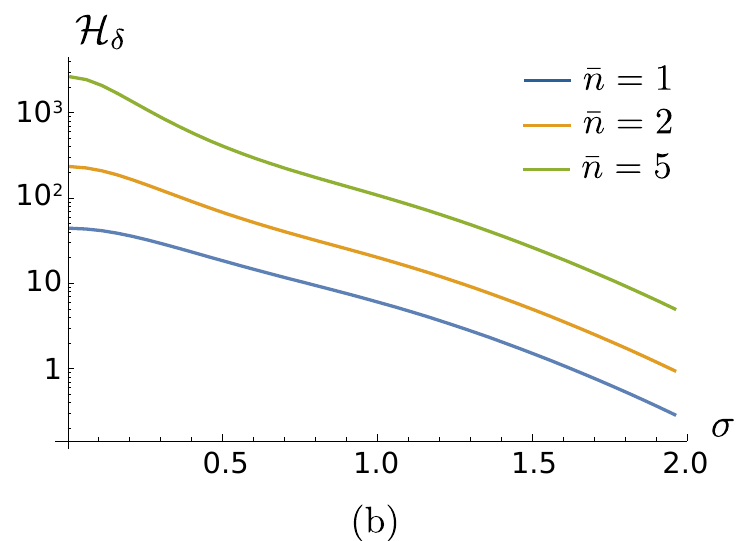}
\end{center}
\caption{Log plots of the dephasing-QFI $\H_{\sigma}$ (a) and nonlinearity-QFI $\H_{\delta}$ (b) for scenario $\caseII$, namely dephasing-Kerr systems, as a function of the phase noise parameter $\sigma=\sqrt{\gamma t}$ for different input coherent state energy $\bar{n}=|\alpha|^2$. Both the QFIs are decreasing functions of the noise $\sigma$, being independent of the nonlinearity parameter $\delta=\kappa t$.}\label{fig:08-QFIDeph}
\end{figure}

\begin{figure}
\begin{center}
\includegraphics[width=0.48\textwidth]{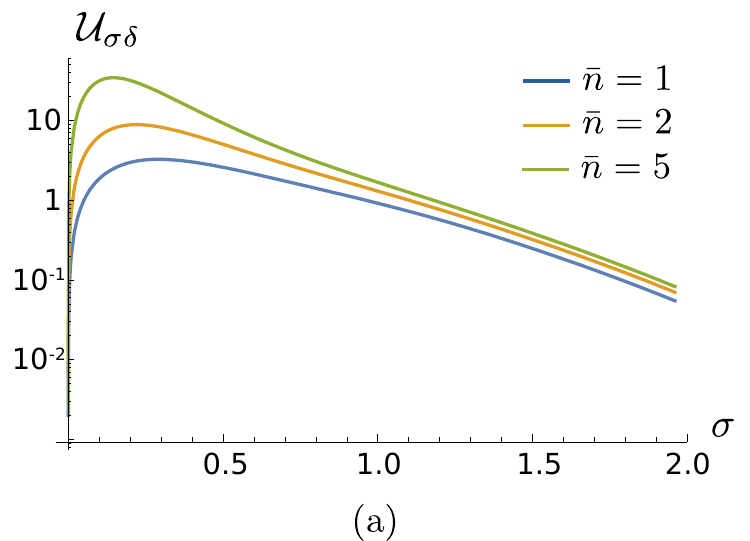} \quad
\includegraphics[width=0.48\textwidth]{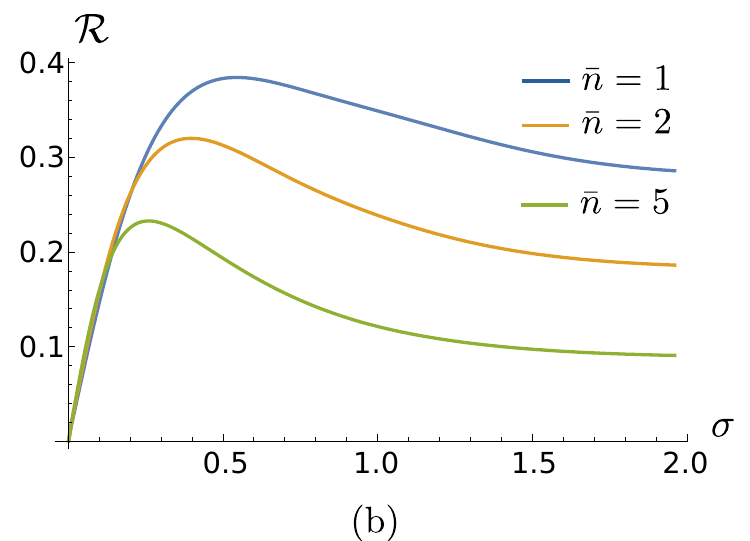}
\end{center}
\caption{(a) Log plot of the Uhlmann quadrature ${\cal U}_{\sigma \delta}$ for scenario $\caseII$, namely dephasing-Kerr systems, as a function of the phase noise parameter $\sigma=\sqrt{\gamma t}$ for different input coherent state energy $\bar{n}=|\alpha|^2$. 
(b) Plot of the quantumness  ${\cal  R}$ for scenario $\caseII$ as a function of $\sigma$ for different $\bar{n}$.
Both the quantities are independent of the nonlinearity parameter $\delta=\kappa t$. In particular, ${\cal  R}$ is a non-monotonic function of $\sigma$, saturating in the large-noise regime to ${\cal  R}_\infty>0$, proving the two parameters to be incompatible.}\label{fig:09-RDeph}
\end{figure}

We numerically compute the QFIM $\QFI= (\H_{\mu\nu})_{\mu,\nu}$, $\mu,\nu=\sigma,\delta$, by Equation~(\ref{eq:QFIM1}), where $\partial_\mu \rho=\sum_{nm}( \partial_\mu \rho_{nm} )|n\rangle \langle m| $, $\mu=\tau,\delta$.
We note that, thanks to the structure of the encoded state, see Equation~(\ref{ew:rhoCOV}), the statistical model is covariant with respect to the nonlinearity, and the corresponding QFIM turns out to be independent of $\delta$ \cite{Paris2009}.
Furthermore, differently from scenario $\caseI$, now the QFIM is a diagonal matrix, that is $\H_{\sigma\delta}=0$, and its diagonal elements $\H_{\sigma}= \H_{\sigma \sigma}$ and $\H_{\delta}= \H_{\delta \delta}$, reported in Figure~\ref{fig:08-QFIDeph}(a) and (b), respectively, are decreasing with the dephasing parameter $\sigma$. In particular, the presence of Kerr susceptibility is irrelevant for the dephasing estimation, as $\H_{\sigma}$ is a function of the sole noise $\sigma$ and input energy $\bar{n}=|\alpha|^2$, in contrast to the case of loss estimation. 
To assess compatibility, we consider the Uhlmann curvature $\UHL$, being an off-diagonal matrix as in~(\ref{eq:UhlmannLoss}), to be computed thanks to~(\ref{eq:Uhl1}). The behaviour is different with respect to  scenario $\caseI$. 
In fact, the off-diagonal term ${\cal U}_{\sigma \delta}$, depicted in Figure~\ref{fig:09-RDeph}(a), is a non-monotonic function of $\sigma$, reaching a maximum at a finite noise $\sigma_{\rm max}$ and, thereafter, decreasing towards $0$. Moreover, like the QFIM, it increases with the signal energy $\bar{n}$, being independent of $\delta$. The non-monotonicity of the Uhlmann quadrature is reflected on the quantumness ${\cal  R}=\sqrt{\det \UHL / \det \QFI}$, plotted in Figure~\ref{fig:09-RDeph}(b). 
For low noise, ${\cal  R}$ increases with $\sigma$ until to reach a maximum, after which it decreases, saturating for $\sigma \gg 1$ to an asymptotic value ${\cal  R}_\infty >0$. The saturation value is lower for increasing energy, whilst the situation is reversed in the low-noise regime, where higher $\bar{n}$ makes $\cal R$ increase.
As a consequence, even in this scenario, the two parameters are not compatible, and cannot be jointly estimated without the addition of an excess noise.

\subsection{Performance of feasible measurements}\label{subsec:FIMcaseII}

\begin{figure}
\begin{center}
\includegraphics[width=0.48\textwidth]{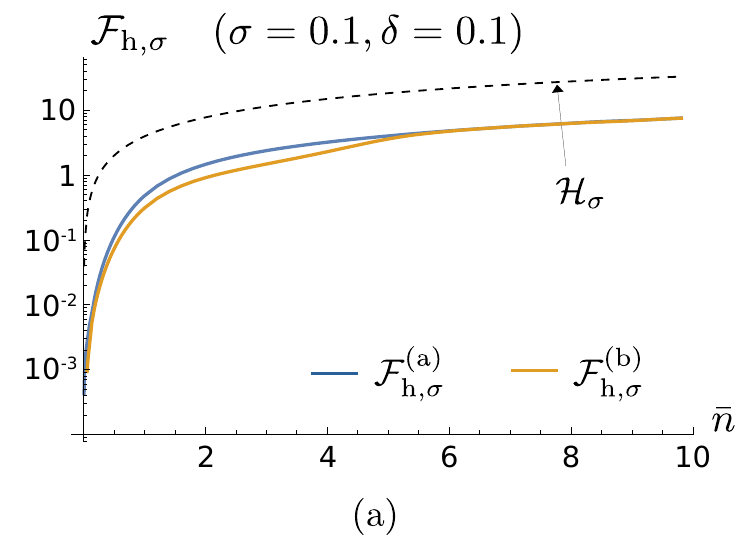} \quad
\includegraphics[width=0.48\textwidth]{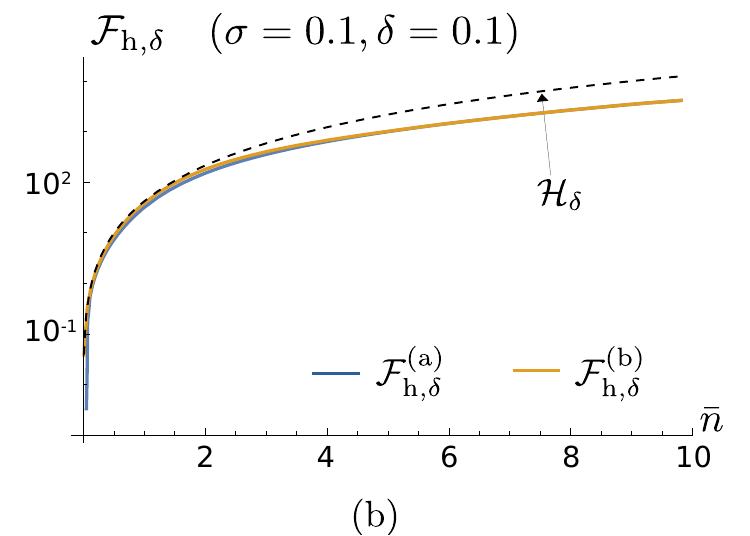}
\end{center}
\caption{Log plot of the dephasing-FI $\F^{(\p)}_{\HD,\sigma}$ (a) and the nonlinearity-FI $\F^{(\p)}_{\HD,\delta}$ (b), $\rm p=a,b$, for scenario $\caseII$, namely dephasing-Kerr systems, as a function of $\bar{n}$. Numerical calculations show that case (c) is almost indistinguishable from case (a). Differently from the lossy-Kerr scenario, now, homodyne detection is strongly suboptimal for the noise estimation, whilst being nearly optimal for the nonlinearity estimation. We set the values $\sigma=0.1$ and $\delta=0.1$ for the dephasing and nonlinearity parameter, respectively.}\label{fig:10-HD-FIDeph-vs-alpha}
\end{figure}

\begin{figure}
\begin{center}
\includegraphics[width=0.55\textwidth]{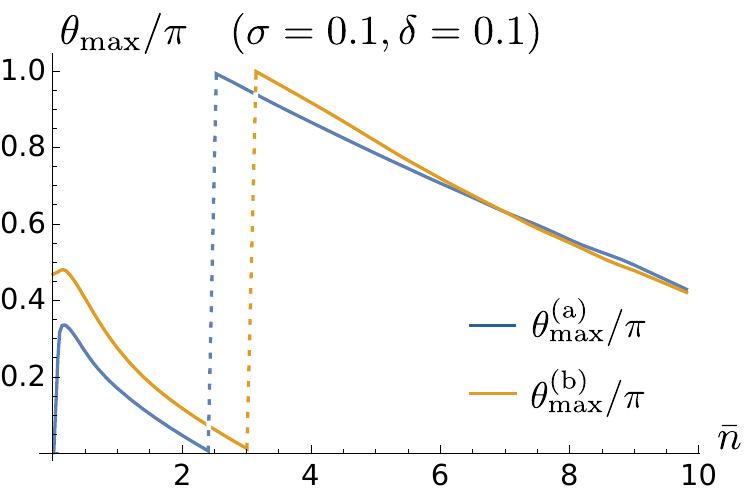}
\end{center}
\caption{Plot of the optimized quadrature phases $\theta_{\rm max}^{(\p)}/\pi$, $\rm p=a,b$, for homodyne detection under scenario $\caseII$, namely dephasing-Kerr systems, as a function of $\bar{n}$. Numerical calculations show that case (c) is almost indistinguishable from case (a).
We set the values $\sigma=0.1$ and $\delta=0.1$ for the dephasing and nonlinearity parameter, respectively.}\label{fig:11-PhiMAXDeph}
\end{figure}

As before, we now quantify the performance of homodyne and DH detection in the joint estimation of $\sigma$ and $\delta$.
Differently from scenario $\caseI$, the physical process associated with phase noise derives from non-Gaussian interaction, therefore we expect Gaussian measurement to be suboptimal even in the limit case $\delta=0$ \cite{Wu2006, Hani2012}.
The homodyne probability $p_\theta(x|\bflambda)$ is retrieved from Equation~(\ref{eq:HomodyneProb}), from which we compute the FIM $\FI_\HD=(\F_{\HD,\mu\nu})_{\mu,\nu}$. As before, we identify the three cases in which we optimize the phase of the measured quadrature to maximize the noise-FI $\F_{\HD,\sigma}= \F_{\HD,\sigma\sigma}$ (a), to maximize the nonlinearity-FI $\F_{\HD,\delta}= \F_{\HD,\delta\delta}$ (b), and to maximize the trace scalar bound ${\cal C}_\HD=1/\Tr[(\FI_{\HD})^{-1}]$ (c), respectively.
Plots of $\F^{(\p)}_{\HD,\sigma}$ and $\F^{(\p)}_{\HD,\delta}$, $\rm p= a,b$, as a function of the input energy $\bar{n}$ are depicted in Figure~\ref{fig:10-HD-FIDeph-vs-alpha}(a) and (b), respectively compared to the corresponding QFIs. Numerical calculations show that case (c) is almost indistinguishable from case (a), therefore we do not explicitly report it in the following Figures.
$\F^{(\a)}_{\HD,\sigma}$ is monotonically increasing with $\bar{n}$, but the separation with respect to $\H_\sigma$ is large, as $\F^{(\a)}_{\HD,\sigma} < 0.25 \H_\sigma$ for all $\bar{n}$, proving homodyne detection to be strongly suboptimal for the noise estimation. On the contrary, the estimation of Kerr nonlinearity is qualitatively similar to that in Section~\ref{subsec:FIMcaseI}: we have $\F^{(\b)}_{\HD,\delta} \gtrsim 0.8 \H_\delta$ for $\bar{n} \lesssim 2.25$, and the homodyne is nearly optimal in the low-energy limit. However, in both the cases, the performances of cases (a) and (b) are close with each other, and almost coincide in the high-energy regime. 
The optimized phase $\theta_{\rm max}^{(\p)}$, $\rm p=a,b$, is depicted in Figure~\ref{fig:11-PhiMAXDeph}, showing a similar behaviour for cases (a) and (b), while the phase jumps are again consequence of the $\pi$-periodicity. 

\begin{figure}
\begin{center}
\includegraphics[width=0.48\textwidth]{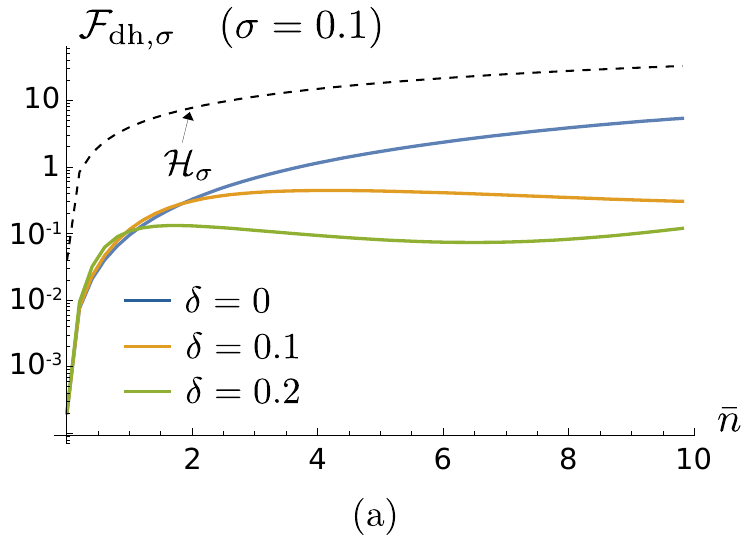} \quad
\includegraphics[width=0.48\textwidth]{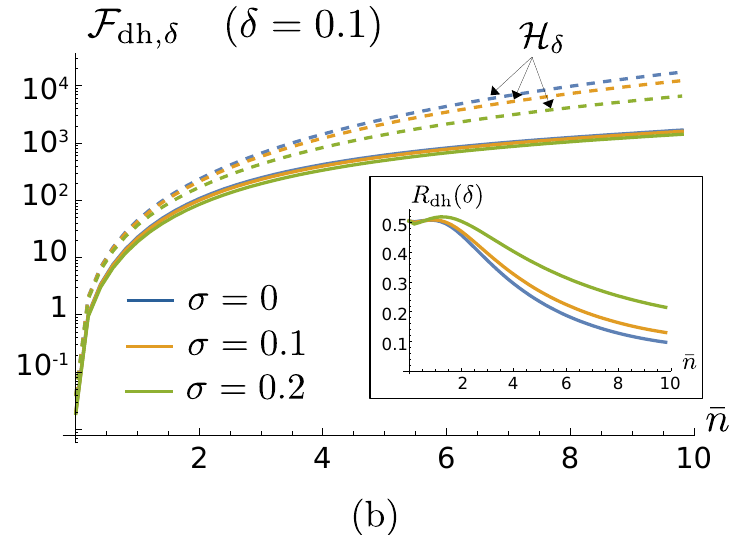}
\end{center}
\caption{(a) Log plot of the dephasing-FI $\F_{\HET,\sigma}$ for scenario $\caseII$, namely dephasing-Kerr systems, as a function of $\bar{n}$ for the dephasing parameter $\sigma=0.1$ and different values of the nonlinearity $\delta$. (b) Log plot of the nonlinearity-FI $\F_{\HET,\delta}$ and plot of the relative ratio $R_\HET(\delta)= \F_{\HET,\delta}/ \H_{\delta}$ (in the inset)  for scenario $\caseII$ as a function of $\bar{n}$ for nonlinearity $\delta=0.1$ and different values of the dephasing $\sigma$.}\label{fig:12-HET-FIDeph-vs-alpha}
\end{figure}

Moving to the case of DH detection, we compute the probability $p_\HET(x,y|\bflambda)$ from Equation~(\ref{eq:DoubleHomodyneProb}), and its corresponding FIM $\FI_\HET=(\F_{\HET,\mu\nu})_{\mu,\nu}$.
The resulting elements $\F_{\HET,\sigma}$ and $\F_{\HET,\delta}$ are plotted in Figure~\ref{fig:12-HET-FIDeph-vs-alpha}(a) and (b), respectively. As regards the noise estimation, we see that DH detection is strongly suboptimal and, in particular, worse than single homodyne detection, since $\F_{\HET,\sigma} <0.2 \H_{\sigma}$. However, the dependence of $\F_{\HET,\sigma}$ on the nonlinearity is nontrivial, as, in the low-energy regime, the noise-FI is enhanced by increasing $\delta$. Instead, the behaviour of the nonlinearity-FI is similar to that in Section~\ref{subsec:FIMcaseI}, that is, DH is suboptimal but $\F_{\HET,\delta}$ is quite robust against the noise, differently than the corresponding QFI, such that the relative ratio $R_\HET(\delta) = \F_{\HET,\delta}/\H_{\delta}$ is increased for larger noise in the high-energy limit.

\begin{figure}
\begin{center}
\includegraphics[width=0.58\textwidth]{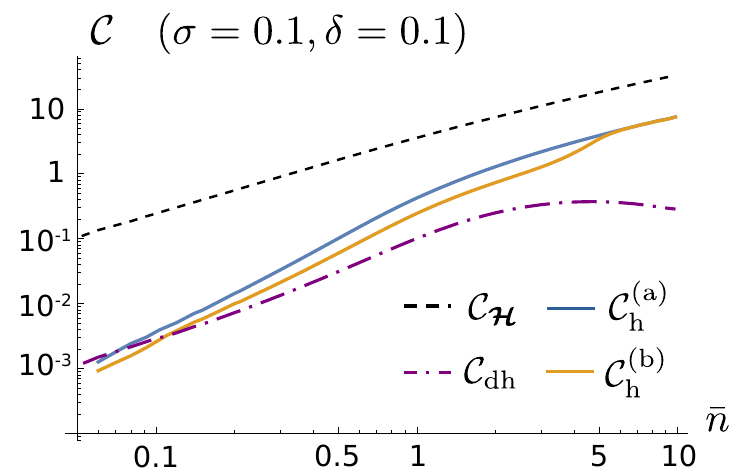}
\end{center}
\caption{Log-log plot of the trace scalar bounds for scenario $\caseII$, namely dephasing-Kerr systems, that is the SLD-QFIM bound ${\cal C}_\QFI= 1/\Tr[\QFI^{-1}]$, and the FIM bounds for homodyne detection, ${\cal C}^{(\p)}_\HD= 1/\Tr[(\FI_\HD^{(\p)})^{-1}]$, $\rm p=a,b$, and DH detection, ${\cal C}_\HET=1/\Tr[\FI_\HET^{-1}]$, as a function of $\bar{n}$. Numerical calculations show that case (c) is almost indistinguishable from case (a). We set the values $\sigma=0.1$ and $\delta=0.1$ for the dephasing and nonlinearity parameter, respectively.}\label{fig:13-TraceDeph}
\end{figure}

Finally, we consider the trace of the inverse FIM matrices as an example of scalar CR bound, depicted in Figure~\ref{fig:13-TraceDeph}.
That is, we compute the quantities ${\cal C}^{(\p)}_{\HD}=1/\Tr[(\FI^{(\p)}_\HD)^{-1}]$, $\rm p=a,b$, ${\cal C}_{\HET}= 1/\Tr[\FI_\HET^{-1}]$, and ${\cal C}_\QFI= 1/\Tr[\QFI^{-1}]$, with the choice of $M=1$ measurements.
As we can see, homodyne detection, in both cases $\rm p=a,b$, is nearly optimal and, differently from scenario $\caseI$, significantly outperforms DH detection for sufficiently high energy. This suggests that probing a single quadrature with optimized phase is worth of implementation in order to achieve a closer performance to the SLD-QFI limit.

\section{Discussion}\label{sec:Resource}

The results obtained in the paper show some relevant qualitative differences between the two scenarios under investigation. In fact, if the behaviour of the nonlinearity estimation is similar in both the scenarios, i.e. decoherence detriments the QFI $\H_\delta$, two opposite situations arise for the decoherence estimation, since the presence of Kerr nonlinearity enhances the loss-QFI $\H_\tau$, while it does not affect the dephasing-QFI $\H_\sigma$.
These considerations raise the question of identifying a proper resource, connected to a quantum property of the encoded states, being somehow responsible for the presence of absence of a QFI enhancement in one scenario or another. This would provide a fundamental interpretation of our results, fostering new methods for engineering more sensitive probe states and measurement schemes.

\begin{figure}
\begin{center}
\includegraphics[width=0.48\textwidth]{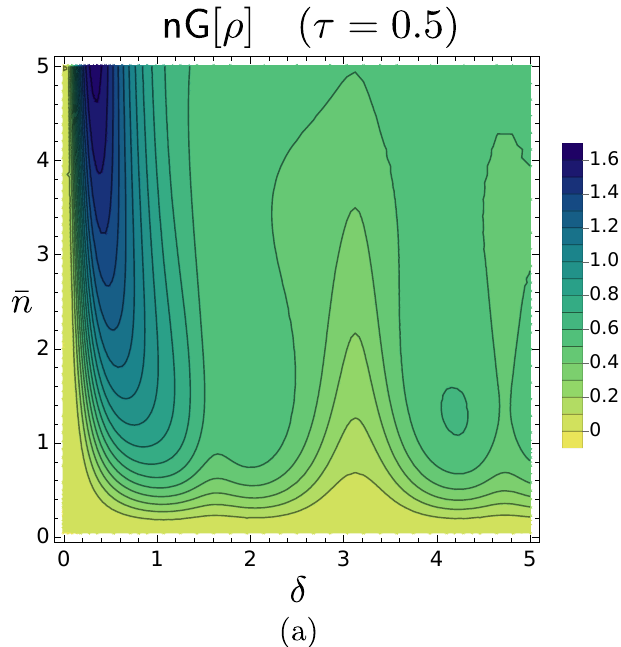} \quad
\includegraphics[width=0.48\textwidth]{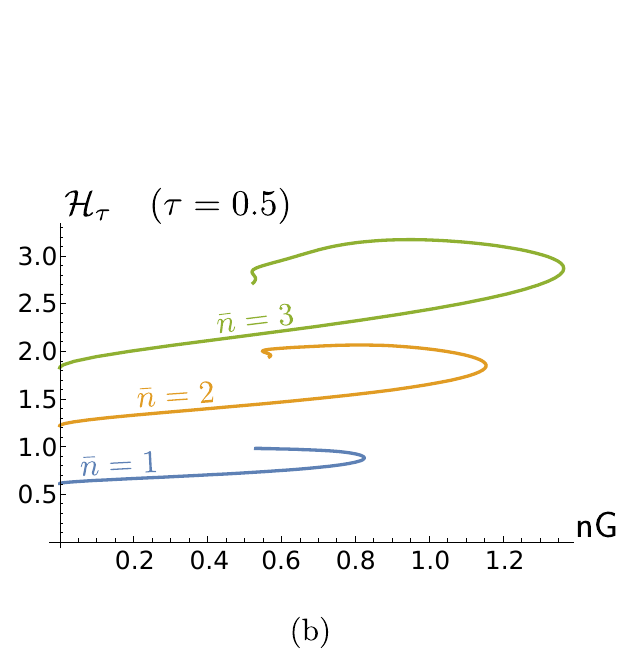}
\end{center}
\caption{
\MN{
(a) Contour plot of the non-Gaussianity ${\sf nG}[\rho]$ for scenario $\caseI$, namely lossy-Kerr systems, as a function of the nonlinearity $\delta$ and the input energy $\bar{n}$ for $\tau=0.5$. ${\sf nG}[\rho]$ is a non-monotonic function of $\delta$ for all $\bar{n}$, increasing in the limit $\delta \ll 1$ and oscillating for $\delta \gg 1$. (b) Parametric plot of the loss-QFI $\H_\tau$ for scenario $\caseI$ as a function of the non-Gaussianity ${\sf nG}[\rho]$, obtained for different $\bar{n}$ (only varying $\delta$) and for $\tau=0.5$. The two functions are not monotone with each other.}
}\label{fig:14-nonG}
\end{figure}

In light of this, non-Gaussianity has been firstly proposed as a suitable candidate \cite{Genoni2009, Rossi2016}, which can be quantified by the difference between the von-Neumann entropy ${\sf S}[\rho]=-\Tr[\rho \log \rho]$ of a quantum state $\rho$ and that of its associated Gaussian state $\rho_{\rm G}$, sharing the same first moment vector and covariance matrix, i.e. ${\sf nG}[\rho]= {\sf S}[\rho_{\rm G}]- {\sf S}[\rho]$ \cite{Navascues2006, Genoni2008}.
In fact, in scenario $\caseI$, namely lossy-Kerr systems, the presence of a nonzero Kerr susceptibility turns a Gaussian statistical model into a non-Gaussian and non-classical one during the evolution. 
\MN{However, as discussed in Section~\ref{sec:LossAndKerr}, the Kerr effect introduces phase-sensitive distortions of the input coherent state in the phase space, therefore the resulting ${\sf nG}[\rho]$, depicted in Figure~\ref{fig:14-nonG}(a), turns out to be a non-monotonic function of $\delta$ for all values of the input energy $\bar{n}$. It increases in the limit $\delta \ll 1$, up to reach a maximum value, whilst, for $\delta \gg 1$, we observe an oscillatory behaviour, reflecting the rotation and stretching effects of the Wigner function of state $\rho$, see Ref.~\onlinecite{Stobinska2008}.
In turn, we have that the maximum amount of non-Gaussianity is achieved at a finite $\delta$, thus qualitatively reproducing the results in Figure~\ref{fig:03-QFILossContour}(a).
Nevertheless, we did not find a quantitative correspondence between the QFI and the non-Gaussianity, as emerges from the parametric plot in Figure~\ref{fig:14-nonG}(b), reporting $\H_\tau$ as a function of ${\sf nG}[\rho]$ for fixed input energy $\bar{n}$ (only varying the nonlinearity $\delta$). As we can see, the two functions are monotone each otehr only for small values of ${\sf nG}$, corresponding to small $\delta$, whereas the highest value of $\H_{\tau}$ is not achieved for the quantum state associated with the highest ${\sf nG}[\rho]$.
}

We reach a similar conclusion by also claiming non-classicality as a resource, described in terms of Wigner negativity and measured, for instance, by the quadrature coherence scale (QCS) \cite{DeBievre2019, Hertz2020, Hertz2023, Griffiet2023}. Indeed, thanks to the nonlinear phase noise effect induced by the Kerr dynamics, the QCS is a non-monotonic function of $\delta$, too.
Furthermore, these arguments fail when applied to scenario $\caseII$, namely dephasing-Kerr systems, where both phase diffusion and Kerr interaction are non-Gaussian, and the non-Gaussianity measure ${\sf nG}[\rho]$ and the QCS are increasing function of both $\sigma$ and $\delta$, whilst $\H_\sigma$ is insensitive to the nonlinearity.

On the other hand, starting from the results of scenario $\caseII$, we may propose as resource the coherence of the quantum statistical model \cite{Baumgratz2014, Winter2016, Xu2016, Tan2017}. In particular, a measure of coherence for a quantum state $\rho$ in a given basis $\{|\varphi_j\rangle\}_j$ has been introduced by Baumgratz {\em et al.} in Ref.~\onlinecite{Baumgratz2014}, and reads ${\sf C}[\rho]= \sum_{j\ne k} |\rho_{jk}|$, with $\rho_{jk}=\langle \varphi_j|\rho|\varphi_k\rangle$.
This captures the behaviour of $\H_\sigma$ in scenario $\caseII$, since the presence of Kerr nonlinearity only appears as a phase in the Fock state expansion of $\rho$, thus not changing its coherence in the Fock basis. However, in scenario $\caseI$, numerical calculations show that nonlinearity reduces the coherence ${\sf C}[\rho]$ with respect to the case $\delta=0$, whilst increasing $\H_\tau$.

In summary, for scenario $\caseI$, namely lossy-Kerr systems, we should conclude that non-Gaussianity in itself (as well as non-classicality) is a necessary but not sufficient condition to enhance the QFI with respect to the performance of coherent probes, since 
there is no monotonic relation between ${\sf nG}[\rho]$ and $\H_\tau$. We note that this conclusion can be further extended to all Gaussian probes, according to recent results derived in \cite{Rossi2016, Albarelli2018, Bressanini2024}. Instead, in scenario $\caseII$, namely dephasing-Kerr systems, quantum coherence seems to be a reasonably necessary condition to improve the precision of dephasing estimation, but we have no evidence that states with higher coherence enhance $\H_\delta$. In both the scenarios, none of the previous figures of merit provide a complete understanding of the sensitivity of the statistical model to the encoded parameters.
\MN{Moreover, the nontrivial interplay between both noise and nonlinearity and the input energy does not allow us to identify a single figure of merit capturing the physics underlying it, thus fostering future studies to face the interesting problem of the resource identification.}

\section{Conclusions}\label{sec:Conc}

\begin{table}[h!]
\caption{\label{Table1} Summary of the results obtained from the QFIM analysis in the two scenarios under investigation, corresponding to loss-Kerr and dephasing-Kerr systems, respectively.}
\vspace{10pt}
\begin{tabular}{@{}l|c|c|c}
\toprule
& Noise estimation & Nonlinearity estimation & Compatibility \\
\hline
Scenario I & enhancement & reduction & no\\
(lossy-Kerr) & & & \\ 
\hline
Scenario II & no enhancement & reduction & no\\
(dephasing-Kerr) & & & \\
\bottomrule
\end{tabular}
\end{table}

\begin{table}[h!]
\caption{\label{Table2} Summary of the results obtained from the FIM analysis in the two scenarios under investigation, corresponding to loss-Kerr and dephasing-Kerr systems, respectively.}
\vspace{10pt}
\begin{tabular}{@{}l|l|c|c|c|c}
\toprule
\multicolumn{2}{c|}{} &\multicolumn{3}{c|}{Homodyne detection} & DH detection\\
\cline{3-5}
\multicolumn{2}{c|}{} & (a) & (b)& (c) &   \\
\hline
& Loss & nearly & strongly & suboptimal & suboptimal \\
& estimation & optimal & suboptimal& & \\ 
\cline{2-6}
Scenario I  & Nonlinearity & suboptimal & nearly & suboptimal & suboptimal \\
(lossy-Kerr) & estimation & & optimal &&   \\
\cline{2-6}
& Scalar bound & suboptimal & strongly & suboptimal & suboptimal \\
& & & suboptimal & & \\
\hline
& Dephasing & strongly & strongly & strongly & strongly \\
& estimation & suboptimal & suboptimal & suboptimal&  suboptimal\\ 
\cline{2-6}
Scenario II & Nonlinearity & suboptimal & nearly & suboptimal & suboptimal \\
(dephasing-Kerr) & estimation & & optimal &&   \\
\cline{2-6}
& Scalar bound & suboptimal & suboptimal & suboptimal & strongly \\
& & & & & suboptimal \\ 
\bottomrule
\end{tabular}
\end{table}

%
%
%

In this paper, we have addressed characterization of noisy Kerr channels, in the presence of either loss or dephasing and a nonzero self-Kerr interaction, by considering a coherent state as a probe. In both the scenarios, we have addressed the joint estimation of the decoherence and nonlinearity parameters in terms of the QFIM, that provides the ultimate bound to the covariance matrix of any unbiased estimator.
In lossy-Kerr systems, we showed that the presence of nonlinearity enhances the loss-QFI $\H_\tau$ in the regime of small loss parameter $\tau$, corresponding to low loss rate of the optical medium or short-distance transmission, and low input energy $\bar{n}$. In particular, we proved that, for any $\bar{n}$, there exists a finite value $\delta_{\rm max}$ of the nonlinearity that maximizes $\H_\tau$, whereas the presence of loss always reduces the QFI for the nonlinearity. Moreover, the Uhlmann curvature is nonzero, thus the two parameters are not compatible and cannot be jointly estimated with maximum precision.
On the other hand, in dephasing-Kerr systems, both the dephasing and nonlinearity QFIs are independent of the nonlinearity parameter $\delta$ and decreasing with the noise amplitude $\sigma$, therefore no enhancement is observed. The Uhlmann quadrature is still nonzero, so the two parameters are incompatible too. A summary of these results is reported in Table~\ref{Table1}.

Thereafter, we considered some relevant examples of feasible POVMs, i.e. homodyne detection and DH detection, and compute the corresponding FIM, to assess their performance with respect to the ultimate bound provided by the QFIM. Table~\ref{Table2} reports a comprehensive sum-up of all the obtained results.
In the presence of lossy-Kerr systems, homodyne detection of a suitably optimized quadrature provides a nearly optimal performance in the low-energy regime, close to the SLD-QCR bound, while DH detection provides a suboptimal solution, although being robust against losses.
Instead, in the dephasing-Kerr scenario, homodyne detection remains nearly optimal only for the nonlinearity estimation, whereas the noise-FI is significantly lower than the corresponding QFI $\H_\sigma$. Similarly, DH detection is strongly suboptimal for the dephasing estimation, whereas it shows robustness also against phase noise in the estimation of the nonlinearity.
Finally, we evaluated a trace scalar bound, proving DH and optimized homodyne detection to be suboptimal in the low- and high- energy regimes, respectively. In particular, in scenario $\caseI$, namely lossy-Kerr systems, DH measurement is closer to homodyne, proving itself as a versatile solution in all conditions.

The obtained results offer a qualitative comparison between the two scenarios and quantify the impact of nonlinearity, being non-negligible in many platforms, for characterization of quantum channels. In particular, they provide a starting point for practical implementations in quantum information protocols, such as sensing \cite{DiCandia2023}, quantum communications \cite{Xiang2018} and quantum key distribution \cite{Lupo2018}.
\MN{Moreover, our analysis provides a first comprehensive study of multiparameter estimation in Kerr media, that may foster future developments by optimization of the input probe states, e.g. by considering squeezed or cat states, that are proved to enhance sensitivity in some particular regimes \cite{Rossi2016}.}

\section*{Acknowledgements}
We acknowledge M.~Frigerio for insightful discussions and F.~Albarelli for useful comments.
This work was done under the auspices of GNFM-INdAM and has been partially supported by EU and MIUR through the project PRIN22-2022T25TR3-RISQUE.
\section*{Conflict of interest}
The authors have no conflicts to disclose.
\section*{Data Availability}
The data that support the findings of this study are available from the corresponding author upon reasonable request.

\end{document}